\documentclass[aps,prl,twocolumn,superscriptaddress]{revtex4-1}
\usepackage{xspace}
\usepackage{bm,physics}
\usepackage{graphicx}
\usepackage{xcolor}
\usepackage{amsmath,amssymb,amsfonts,amsthm}
\usepackage[colorlinks=true,linkcolor=blue,anchorcolor=red,citecolor=blue,urlcolor=blue]{hyperref}
\usepackage{multirow}
\usepackage{diagbox}

\def \T {\mathcal{T}}

\def \P {\mathcal{P}}

\def \UU {\mathcal{U}}
\def \Q {\mathcal{Q}}
\def \S {\mathcal{S}}

\def \k {\bm{k}}
\def \Z {\mathbb{Z}}

\def \H {\mathcal{H}}

\begin{document}

\title{Topological classification for chiral symmetry with non-equal sublattices }

\author{J. X. Dai}
\affiliation{Department of Physics and HK Institute of Quantum Science \& Technology, The University of Hong Kong, Pokfulam Road, Hong Kong, China}

\author{Y. X. Zhao}
\email[]{yuxinphy@hku.hk}
\affiliation{Department of Physics and HK Institute of Quantum Science \& Technology, The University of Hong Kong, Pokfulam Road, Hong Kong, China}

\begin{abstract}
	Chiral symmetry on bipartite lattices with different numbers of $A$-sites and $B$-sites is exceptional in condensed matter, as it gives rise to zero-energy flat bands. Crystalline systems featuring chiral symmetry with non-equal sublattices include Lieb lattices, dice lattices, and particularly Moiré systems, where interaction converts the flat bands into fascinating many-body phases.
	In this work, we present a comprehensive classification theory for chiral symmetry with non-equal sublattices. First, we identify the classifying spaces as Stiefel manifolds and derive the topological classification table. Then, we extend the symmetry by taking $\mathcal{PT}$ symmetry into account, and ultimately obtain three symmetry classes corresponding to complex, real, and quaternionic Stiefel manifolds, respectively. Finally, we apply our theory to clarify the topological invariant for $\mathcal{PT}$-invariant Moiré systems and construct physical models with Lieb and dice lattice structures to demonstrate our theory. Our work establishes the theoretical foundation of topological phases protected by chiral symmetries with non-equal sublattices.
\end{abstract}

\maketitle

\begin{figure}
	\centering
	\includegraphics[width=\columnwidth]{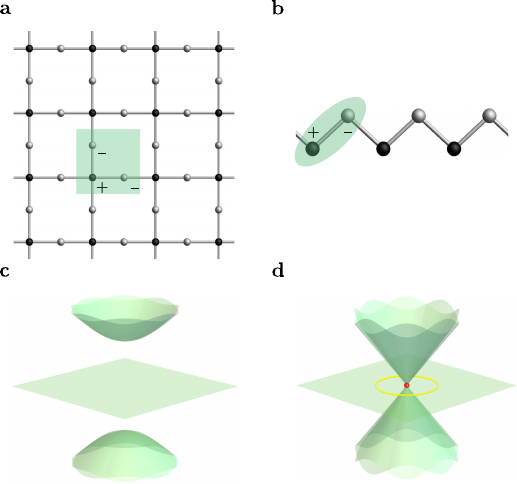}
	\caption{\textbf{a} and \textbf{b} illustrate the Lieb lattice model and the Su-Schrieffer-Heeger model, respectively. Unit cells for them are marked by shadowed regions, where $A$-sites and $B$-sites are marked in black and grey and assigned signs $\pm$ by the sublattice symmetry, respectively.  \textbf{c} and \textbf{d} exhibit typical gapped and nodal-point band structures under CSNES, respectively.  In \textbf{d}, topological charge is defined on the yellow loop surrounding the nodal point. } \label{FIG1}
\end{figure}

{\color{blue}\textit{Introduction}} Sublattice on bipartite lattices with non-equal sublattices is an exceptional symmetry in condensed matter. This symmetry concerns a bipartite lattice hosting different numbers of $A$-sites and $B$-sites for the two sublattices~\cite{regnault2022catalogue,cualuguaru2022general} and anti-commutes with the Hamiltonian that only couples A-sites and B-sites. Such sublattice symmetry is referred to as chiral symmetry with non-equal sublattices (CSNES). Elementary examples include the Lieb lattices and dice lattices \cite{taie2020spatial,slot2017experimental,Vicencio2015PRL,Muk2015PRL,Xia2018PRL,Julku2016PRL,Andrj2015PRA,mohanta2023majorana,Wang2011PRB}, where every unit cell comprises a single $A$-site and two $B$-sites, as illustrated by Fig.~\ref{FIG1}\textbf{a}. 

A characteristic of CSNES is the presence of $|M-N|$ flat bands at zero energy, where $M$ and $N$ are $A$ and $B$ site numbers in a unit cell, respectively (see Fig.~\ref{FIG1}\textbf{c} and \textbf{d}). The flat bands resulting from CSNES have recently attracted significant interest in Moir\'{e} systems, such as twisted bilayer graphene, where fascinating quantum phases can be formed through interactions~\cite{regnault2022catalogue,cualuguaru2022general,Vishwanath2019PRB}. Additionally, the Fermi points feature Berry dipole charge, as recently revealed~\cite{Hopf_dipole,Berry-dipole_Shuang}.

Naturally, a fundamental question arises: How to derive the topological classifications for CSNES.  

Conventionally, chiral symmetry is considered under the framework of the tenfold Altland-Zirnbauer symmetry classes (AZ classes) ~\cite{Schnyder2008PRB,Atiyah-KR,Kitaev2009AIP,AZ1997PRB,ZhaoYXWang13prl}. Various topological classifications for the tenfold AZ classes play a seminal role in the development of symmetry protected topological phases that have dominated modern condensed matter physics and beyond in recent years \cite{Volovik:book,SCZHANGRMP2011,Shun-Qing-Shen,KaneRMP2010,SchnyderRMP2016,Teo2010PRB}.  In applying the tenfold topological classifications to bipartite lattices, there is a presumption, i.e., in each unit cell the  number of $A$-sites is equal to that of $B$-sites, as exemplified by the Su-Schrieffer-Heeger model in Fig.~\ref{FIG1}\textbf{b}. However, this presumption has been violated in the case of CSNES. Consequently, the chiral AZ classes always assume a gap at zero energy, which is inconsistent with the aforementioned zero-energy flat bands originating from CSNES.



This work is devoted to presenting a conprehentive classification theory for CSNES. The key step is to identify the classifying space for a CSNES. Our analysis shows that the classifying space is the Stiefel manifold $V_M(\mathbb{C}^N)$, namely the space consisting of all possible sets of $M$ orthonormal vectors in an $N$-dimensional complex vector space \cite{TopologySM,Wiki}. Here, we assume $N\ge M$ without loss of generality.  

As the CSNES can be regarded as a generalization of class AIII in the AZ classes, we further consider spacetime inversion symmetry $\mathcal{PT}$ to extend CSNES~\cite{WangPRL2020,YXZHAOPRL2016,YXZHAOPRL2017,BJYPRL2018,BJYPRX2019,Tomas2017PRB,xue2023stiefel,DAI2021PRB,ShengXLPRL2019}. 
We identify totally three symmetry classes with the classifying spaces $V_N(\mathbb{C}^N)$, $V_N(\mathbb{R}^N)$ and $V_N(\mathbb{H}^N)$, resembling Dyson's threefold way~\cite{dyson}. The topological classifications for both gapped and gapless systems are given by the homotopy groups $\pi_d[V_M(\mathbb{F}^N)]$ with $\mathbb{F}=\mathbb{C}$, $\mathbb{R}$ and $\mathbb{H}$. For gapped systems, $d$ is just the spatial dimensionality, while for gapless systems $d$ corresponds to the spatial codimension of the gapless modes~\cite{Volovik:book,ZhaoYXWang13prl,Co-dim} ( Fig.~\ref{FIG1}\textbf{d} ).

As an important application, the classification theory is applied to clarify the topological invariant of the Po-Zou-Senthil-Vishwanath (PZSV) model, which captures the essential physics of various Mori\'{e} systems~\cite{cualuguaru2022general,Vishwanath2019PRB}. We further demonstrate our theory by constructing a 3D topological insulator on the Lieb lattice model corresponding to the unit element of $\pi_3[V_{1}(\mathbb C^2)]\cong\mathbb{Z}$, a 1D topological insulator and a 2D topological semimetal on dice lattices corresponding to the nontrivial element of $\pi_1[V_1(\mathbb R^2)]\cong\mathbb{Z}_2$. 

{\color{blue}\textit{The general sublattice symmetry}} Let us start with formulating the sublattice symmetry. On a bipartite lattice, the sublattice symmetry is essentially a $\mathbb{Z}_2$ gauge transformation:
\begin{equation} \label{eq:G_transf}
S: |i\rangle\mapsto (-1)^{s(i)} |i\rangle,
\end{equation}
where $s(i)=0$ or $1$ for $i$ being a $A$-site or $B$-site, respectively. Let $M$ and $N$ be the site numbers of $A$-sites and $B$-sites in each unit cell. The sublattice symmetry operator $S$ is represented in momentum space as
\begin{equation}
	\mathcal{S}=\begin{bmatrix}
		1_M & \\
		 & -1_N
	\end{bmatrix}
\end{equation}
with $1_n$ being the $n\times n$ identity matrix.  

A real-space Hamiltonian $H$ preserves the sublattice symmetry if it solely consists of hopping terms  between the two sublattices, i.e., $H=\sum_{ij}t_{ij}|i\rangle\langle j|$ with $i\in A$ and $j\in B$ for each $t_{ij}|i\rangle\langle j|$. It is clear that the gauge transformation \eqref{eq:G_transf} inverses every hopping term $t_{ij}\mapsto -t_{ij}$, and therefore inverses the Hamiltonian $H$. Accordingly, the momentum-space Hamiltonian $\mathcal{H}(\bm{k})$ satisfies $\mathcal{S}\mathcal{H}(\bm{k})\mathcal{S}^\dagger=-\mathcal{H}(\bm{k})$, which may be more elegantly expressed as the anti-commutation relation,
\begin{equation}\label{eq:anti-comm}
	\{ \mathcal{H}(\bm{k}),\mathcal{S}\}=0.
\end{equation}
The anti-commutation relation \eqref{eq:anti-comm} requires the Hamiltonian be in the anti-diagonal form,
\begin{equation}\label{eq:anti-diagonal}
	\H(\k)=\begin{bmatrix}
		&\Q(\k)\\\Q^\dagger(\k)&
	\end{bmatrix},
\end{equation}
where $\Q(\k)$ is an $M \times N$ matrix and therefore $\Q^\dagger(\k):=[\Q(\k)]^\dagger$ an $N \times M$ matrix.

The spectrum of $\mathcal{H}(\bm{k})$ in the anti-diagonal form is determined by the singular values of $\Q(\bm{k})$. To see this, we first perform the singular value decomposition
\begin{equation}\label{eq:SVD}
	\Q(\k)=\mathcal V(\k)\mathcal{D}(\k)\UU^\dagger(\k).
\end{equation}
Here, $\mathcal{V}$ and $\mathcal{U}$ are square matrices with rank $M$ and $N$, respectively. $\mathcal{D}$ is an $M\times N$ matrix of the form $\mathcal{D}=[\Lambda,\mathbf{0}]$, where $\Lambda$ is a diagonal $M\times M$ square matrix and $\mathbf{0}$ the zero $M\times(N-M)$ matrix. The diagonal entries of $\Lambda$ are just the singular values $\lambda_i$ of $\Q$ with $\lambda_i\ge 0$. Then, the Hamiltonian $\H(\k)$ can be expressed in the decomposition form,
\begin{equation}\label{Eq:Uni-trans}
	\H(\bm{k})=\begin{bmatrix}
		\mathcal V&\\&\UU
	\end{bmatrix}\begin{bmatrix}
		&\mathcal{D}(\bm{k})\\\mathcal{D}^\dagger(\bm{k})&
	\end{bmatrix}\begin{bmatrix}
		\mathcal V^\dagger&\\&\UU^\dagger
	\end{bmatrix}.
\end{equation}
Then, the spectrum of $\mathcal{H}(\bm{k})$ can be read off as 
\begin{equation}
 \lambda_1(\bm{k}),-\lambda_1(\bm{k})\cdots, \lambda_M(\bm{k}),-\lambda_M(\bm{k}), 0,\cdots 0,
\end{equation}
with $N-M$ zeros. Each single value $\lambda_i$ contributes a pair of eigen energies $\pm\lambda_i$. Thus, we observe a feature of the CSNES, i.e., generically there exist $|N-M|$ flat bands at the zero energy.

{\color{blue}\textit{The classifying spaces}} We then proceed to derive the classifying spaces. For the tenfold symmetry classes, this is done by considering a one-point symmetry-preserving Hamiltonian, where an energy gap is assumed at zero energy and eigen energies are flattened to be $\pm 1$. Here, with the $|N-M|$ zero modes for the one-point Hamiltonian $\mathcal{H}$ with the CSNES, we assume two energy gaps below and above the zero energy and flatten positive and negative eigen energies as $\pm 1$, respectively. 
\begin{table}[t]
	\begin{tabular}{c|c|c| c c c}
		$(\P\T)^2$ & $V_M(\mathbb{K}^N)$ & $N-M$ & $d=1$ & $d=2$  & $d=3$\\
		\hline
		\hline
		0  &  ~$V_M(\mathbb C^{N})$~  & 0  & $\Z$&0&$\Z$\\
		& &1 & 0&0&$\Z$\\
		\hline
		1  & $V_M(\mathbb R^{N})$   &0 & $\Z_2$&0&$\Z$\\
		& &1  & $\Z_2$&0&$\Z$\\
		& &2 & 0&$2\Z$&$\Z$\\
		& &3  & 0&0&$\Z_2$\\
		\hline
		-1 & $V_M(\mathbb H^{N})$  &0 & $0$&0&$\Z$
	\end{tabular}
	\caption{Topological classification table of the Stiefel manifolds. The first column specifies the three cases of $PT$ symmetry, which determines the classifying space $V_M(\mathbb{K}^N)$. Here, $(\P\T)^2=0$ indicates the absence of $\P\T$ symmetry. For each classifying space, $N-M$ is fixed at various values while $N$ and $M$ are presumed to be sufficiently large. The corresponding homotopy groups $\pi_d(V_M(\mathbb{K}^N))$ are presented for $d=1,2,3$. Note that $\pi_d(V_M(\mathbb{K}^N))=0$ if $N-M$ is greater than the exhibited range.}\label{Table:Realtable}
\end{table}

The assumption of two energy gaps means all singular values of $\Q$ are nonzero, and the flattening of the spectrum of $\mathcal{H}$ is equivalent to the flattening of all singular values of $\Q$ to $1$. Let $\widetilde{\mathcal{O}}$ denote the flattened version of any operator $\mathcal{O}$. Then, $\widetilde{\Lambda}=1_{M}$, and $\widetilde{\mathcal{D}}=(1_M, 0_{M\times(N-M)})$. Consequently,
\begin{equation}\label{eq:SimplifiedSVD1}
	\begin{split}
			\widetilde{\Q}&=\mathcal V[1_M, \mathbf{0}]\UU^\dagger=[1_M, \mathbf{0}]\begin{bmatrix}
			\mathcal V &\\ & 1_{N-M}
		\end{bmatrix}\UU^\dagger.
	\end{split}
\end{equation}
Hence, any $\Q$ can be flattened to the form,
\begin{equation}\label{eq:canonical_form}
	\widetilde\Q=[1_M, \mathbf{0}]\mathcal{W},
\end{equation}
where $\mathcal{W}$ can be any $N\times N$ unitary matrix, namely $\mathcal W\in \mathrm{U}(N)$.

Let us look into the geometric meaning of Eq.~\eqref{eq:canonical_form}. The $i$th row of $\mathcal{W}$ is a vector $v_i$ in the space $\mathbb{C}^N$, and all the $N$ vectors $v_i$ form an orthonormal basis of $\mathbb{C}^N$, i.e., $(v_i,v_j)=\delta_{ij}$. The multiplication by $[1_M, \mathbf{0}]$ on the left side of $\mathcal{W}$ just selects the first $M$ rows, namely the first $M$ vectors $v_i$ with $i=1,2,\cdots,M$. Since $\mathcal{W}$ can exhaust all possible orthonormal bases of $\mathbb{C}^N$, the space of $\Q$ is just the space of all possible $M$ orthonormal vectors in $\mathbb{C}^N$, which is just the definition of the Stiefel manifold denoted by $V_M(\mathbb{C}^N)$. Thus, the classifying space of the CSNES is just the complex Stiefel manifold $V_M(\mathbb{C}^N)$.

It is significant to notice that $\mathcal{W}$ is not uniquely determined by $\widetilde{\Q}$, because $\widetilde{\Q}$ is invariant under the transformation,
\begin{equation}\label{Eq:Gauge}
	\mathcal W\mapsto \begin{bmatrix}
		1_M &\\ & \mathcal{G}
	\end{bmatrix}\mathcal W,
\end{equation}
for any unitary matrix $\mathcal{G}\in \mathrm U(N-M)$. The inverse is also true. If $\mathcal{W}$ and $\mathcal{W}'$ corresponds to the same $\widetilde{\Q}$, there exists a unitary matrix $\mathcal{G}\in \mathrm{U}(N-M)$ with $\mathcal{W}'=\mathrm{diag}[1_M,\mathcal{G}]\mathcal{W}$. Thus, we obtain the representation of the complex Stiefel manifold,
\begin{equation}\label{Eq:Complexmanifold}
	V_M(\mathbb C^N)=\mathrm{U}(N)/\mathrm{U}(N-M).
\end{equation}
The special case of $V_N(\mathbb C^N)=\mathrm{U}(N)$ agrees with the classifying space of class AIII in the tenfold symmetry classes.

The topological classifications are given by the homotopy groups $\pi_d[V_M(\mathbb C^N)]$ as aforementioned. For a fixed value of $N-M$ and a given $d$, $\pi_d[V_M(\mathbb C^N)]$ will stabilize to a fixed group if $M$ surpasses a threshold value. This is the so-called the stability of homotopy groups. The stable homotopy groups for physical dimensions are tabulated in Tab.~\ref{Table:Realtable}. See the SM for derivation details \cite{Supp}.

{\color{blue}\textit{The threefold way of classifying spaces}}  We then include  $\mathcal{PT}$ symmetry into consideration to generalize the five chiral AZ classes with ordinary sublattice symmetry.
Besides class AIII with only the chiral symmery, the other four are specified by $(\mathcal{PT})^2=\pm 1$ and  $\mathcal{PT}\mathcal{S}=\pm\mathcal{S}\mathcal{PT}$. For the later, which commutation relation is satisfied is determined by whether inversion exchanges the two sublattices. But, for CSNES, the two sublattices cannot be exchanged, since they have different site numbers in each unit cell. Then, only the commutation relation $\mathcal{PT}\mathcal{S}=\mathcal{S}\mathcal{PT}$ is allowed. Then, there exist only two possibilities, namely $(\mathcal{PT})^2=\pm 1$.

For $(\mathcal{PT})^2=1$, the operator can always be represented as $\mathcal{PT}=\mathcal{K}$ with $\mathcal{K}$ the complex conjugation. With $\mathcal{PT}=\mathcal{K}$, $\mathcal{H}$ becomes a real symmetric matrix, and the Hilbert space is essentially a real space $\mathbb{R}^N$. Accordingly, the classifying space is the real Stiefel manifold,
\begin{equation}\label{Eq:Realmanifold}
	V_M(\mathbb R^N)=\mathrm{O}(N)/\mathrm{O}(N-M).
\end{equation}

For $(\mathcal{PT})^2=-1$, each energy band bears the Kramers degeneracy, and the $\mathcal{PT}$ operator can be represented as $\mathcal{PT}=-i\sigma_2\mathcal{K}$ with the Pauli matrix $\sigma_2$ acting in the twofold degenerate space. $\mathcal{PT}=-i\sigma_2\mathcal{K}$ together with the imaginary unit $i$ form generators of the quaternionic field $\mathbb{H}$. Hence, the classifying space is the quaternionic Stiefel manifold,
\begin{equation}\label{Eq:Quaternionicmanifold}
	V_M(\mathbb H^N)=\mathrm{Sp}(N)/\mathrm{Sp}(N-M).
\end{equation}

The stable homotopy groups of real and quaternionic Stiefel manifolds are tabulated in Tab.~\ref{Table:Realtable}, and the derivation details are provided in the SM \cite{Supp}. The three classifying spaces resemble Dyson's threefold way for symmetry classes~\cite{dyson}.

{\color{blue}\textit{Application and demonstration}} The PZSV model for various Mori\'{e} systems features CSNES that leads to the zero-energy flat bands and $\mathcal{PT}$ symmetry for the real band structure. Therefore, the model well fits into our framework. The classifying space is $V_M(\mathbb{R}^N)$, and the corresponding homotopy group is $\pi_2(V_M[\mathbb{R}^N)]\cong 2\mathbb{Z}$ with $N-M=2$. The detailed analysis of the model is quite technical and can be found in the SM~\cite{Supp}. The result is that the model does represent nontrivial topology of the classification.

We now present models for simple and clear demonstration of the topological classifications in Tab.~\ref{Table:Realtable}.

\textit{3D topological insulator} According to Tab.~\ref{Table:Realtable}, the homotopy group $\pi_3[V_{N-1}(\mathbb C^N)]\cong\mathbb{Z}$ means there exist $3$D chiral symmetric topological insulators with $A$-sites one more than $B$-sites in each unit cell.

\begin{figure}[t]
	\centering
	\includegraphics[width=\columnwidth]{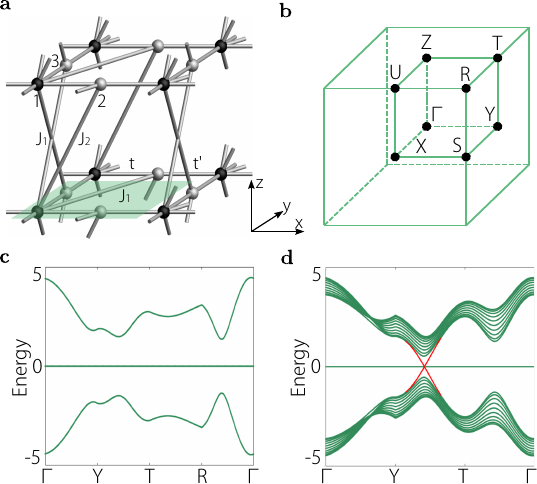}
	\caption{\textbf{a} The $3$D lattice model as a stack of layers of Lieb lattices. Each unit cell contains one $A$-site and two $B$-sites marked in black and grey, respectively. \textbf{b} The Brillouin zone (BZ) with high-symmetry points marked. \textbf{c} The bulk band structure along the path specified by the high symmetry points. \textbf{d} The band structure with a boundary opened normal to the $x$ direction. It is plotted along the path specified by the high symmetry points in the boundary BZ. The boundary states are marked in red.  We set $t=0.5$, $t'=1$, $J_1=1$ and $J_2=\text i$.} \label{FIG2}
\end{figure}

Such a model is constructed on the $3$D lattice as a stack of layers of Lieb lattices, where each unit cell contains one $A$-site and two $B$-sites  as illustrated in Fig.~\ref{FIG2}\textbf a.  
The Hamiltonian is given by
\begin{equation}\label{Eq:C21model}
\H_{\text{3D}}(\k)=\begin{bmatrix}
0&w_1(\k)&w_2(\k)\\w^*_1(\k)&0&0\\ w^*_2(\k)&0&0
\end{bmatrix},
\end{equation}
where $w_1(\k)=t+t'e^{\text ik_x}+J_1e^{-\text ik_y}+J_2e^{-\text ik_z}$ and $w_2(\k)=t+t'e^{\text ik_y}+2J_1\cos k_z$.  

Since hopping occurs only between different sublattices, the CSNES, represented by 
$\S=\mathrm{diag}[1, -1_2]$, is preserved. The energy spectrum of $\H_{\text{3D}}(\k)$ can be derived as $\{0,\lambda(\k),-\lambda(\k)\}$ with $\lambda(\k)=(|w_1|^2+|w_2|^2)^{1/2}$. We observe that the band structure features two gaps separated by a flat band at the zero energy for the parameters given in the caption of Fig.~\ref{FIG2}\textbf c.

From the derivation of the topological classification in the SM \cite{Supp}, we learn that the topological invariant $\mathcal{N}$ is just the winding number of $\mathcal W(\k)$ defined in Eq.~\eqref{eq:canonical_form}, i.e.,
\begin{equation}\label{Eq:C21topo}
	\mathcal{N}=\frac{1}{24\pi^2}\int_{\mathrm{BZ}} d^3k~\epsilon^{ijk}\mathrm{tr}~\mathcal W\partial_i\mathcal W^\dagger\mathcal W\partial_j\mathcal W^\dagger\mathcal W\partial_k\mathcal W^\dagger,
\end{equation}
where $\epsilon^{ijk}$ is the completely antisymmetric tensor. 

Particularly, for our model \eqref{Eq:C21model},
$
	\mathcal W(\k)=d_0\sigma_0-\text{i}\sum_{i=1}^3d_i\sigma_i,
$
where $d_0=\Re(w_1/\lambda)$, $d_1=\Im(w_2/\lambda)$, $d_2=\Re(w_2/\lambda)$, $d_3=\Im(w_1/\lambda)$, $\sigma_0=1_2$ and $\sigma_i$ are the Pauli matrices. Then, $d_\mu$ is just a vector on the unit sphere $S^{3}$, and $ \mathcal{N}$ is just the winding number of the unit vector $d_\mu$ wrapping over the Brillouin zone, i.e., 
$
	\mathcal{N}=\frac{1}{2\pi^2}\int_{T^3}d^3k~\epsilon^{\mu\nu\rho\lambda}
	d_\mu\partial_{k_1}d_\nu\partial_{k_2}d_\rho\partial_{k_3}d_\lambda.
$
For the parameter values in Fig.~\ref{FIG2}\textbf c, we obtain $\mathcal{N}=-1$. 

The nontrivial topological invariant leads to boundary bands across the two energy gaps, which connect the states in the $3$D bulk bands. This is shown in Fig.~\ref{FIG2}\textbf d, where the boundary is opened perpendicular to the $x$ direction.

\begin{figure}
	\centering
	\includegraphics[width=\columnwidth]{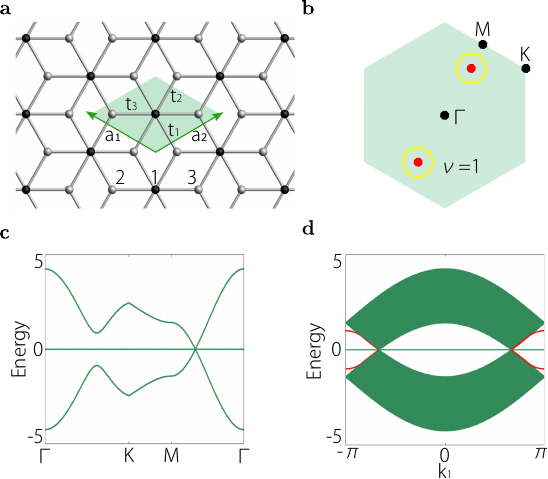}
	\caption{\textbf{a} The dice lattice structure. $A$-sites and $B$-sites are marked in black and grey, respectively, and the unit cell is indicated by the shadowed parallelogram.   $\mathbf{a}_1$ and $\mathbf{a}_2$ denote the two unit lattice vectors. \textbf{b} The Brillouin zone. The two nodal points are highlighted as two red dots, and high symmetry points are marked as black dots. \textbf{c} The bulk band structure along the $\Gamma-K-M-\Gamma$ line in \textbf{b}. \textbf{d} The band structure with a boundary opened parallel to $\bm{a}_1$ direction. The edge Fermi arcs are marked in red. We set $t_1=t_2=t_3=1$. } \label{FIG3}
\end{figure}

\textit{$2$D topological semimetal with $PT$ symmetry}
The homotopy group $\pi_1[V_{N-1}(\mathbb{R}^N)]\cong \mathbb{Z}_2$ in Tab.~\ref{Table:Realtable} indicates that there exist $1$D topological insulator, $2$D topological nodal point and $3$D topological nodal line, protected by CSNES and $\mathcal{PT}$  symmetry without spin-orbit coupling. A model for the $1$D topological insulator can be found in the SM \cite{Supp}, and we present a $2$D topological semimetal on the dice lattice here.

As illustrated in Fig.~\ref{FIG3}\textbf a, each unit cell contains one $A$-site and two $B$-sites, and symmetry operators are represented as $\S=\mathrm{diag}[1,-1_2]$ and $	\mathcal{PT}=\mathrm{diag}[1,\sigma_1]\mathcal{K}$.
The $\mathcal{PT}$ symmetry is in the form because the inversion symmetry is centered at the $A$-site and exchanges two $B$-sites. The Hamiltonian is given by
\begin{equation}
	\H_\text{2D}(\k)=w(\k)S_{+}+w^*(\k)S_{-},
\end{equation}
where $w(\k)=t_3+t_1e^{-\text i\k\cdot \bm{a}_1}+t_2e^{\text i \k\cdot \bm{a}_2}$. Here, $S_{\pm}=(S_{x}\pm \mathrm{i} S_{y})/\sqrt{2}$, where $S_i$ are the spin-$1$ operators with the explicit matrix representation given in the SM \cite{Supp}. Hamiltonians in this form are called Maxwell Hamiltonians \cite{ZHUYANQING2017PRA,TAN2018PRL,Stone2016IJMPB}. Notably, the Hamiltonian resembles that of graphene with $\sigma_{\pm}$ replaced by $S_{\pm}$. 


The energy spectrum is derived as $0$, $\pm \sqrt{2}|w(\k)|$, resembling that of graphene with an additional flat band at zero energy. Hence, it generically corresponds to a semimetal with two nodal points at $\pm \bm{K}$ [see Fig.~\ref{FIG3}\textbf b], each now featuring threefold degeneracy.  The $k\cdot p$ model at $\bm{K}$ is given by
\begin{equation}
	h(\bm{q})=vq_{+}S_{+}+vq_{-}S_{-}
\end{equation}
with $q_{\pm }=(v_{x}q_x\pm \mathrm{i} v_yq_y)/v$. Here, $\bm{q}=\bm{k}-\bm{K}$. The parameters $v_x$ and $v_y$ are determined by $t_i$, and $v=(v_x^2+v_y^2)^{1/2}$. The $k\cdot p$ model at $-\bm{K}$ is related to $h(\bm{q})$ by time-reversal symmetry.

To formulate the topological charge, we first perform a unitary transformation $U=\mathrm{diag}[1,e^{-i\pi/4}e^{i\pi\sigma_1/4}]$ with $U\mathcal{PT}U^\dagger=\mathcal{K}$. Then, we obtain the real Hamiltonian $\mathcal{H}_\text{2D}^{\mathbb{R}}(\k) = U \mathcal{H}_\text{2D}(\k) U^\dagger=\sqrt{2}w_{+}(\k)S_{+}+\sqrt{2}w_{-}(\k)S_{-}$ with $w_{\pm}=(\mathrm{Re}w\pm \mathrm{Im}w)/\sqrt{2}$. The topological charge of each nodal point is formulated on a circle $S^1$ enclosing $\bm{K}$ [see Fig.~\ref{FIG3}\textbf b]. Since $\mathcal{H}_\text{2D}^{\mathbb{R}}(\phi)$ with $\phi\in S^1$ is gapped, we can derive $\mathcal W(\phi)\in \mathrm{O}(2)$ defined in Eq.~\eqref{eq:canonical_form} from $\mathcal{H}_\text{2D}^{\mathbb{R}}(\phi)$. The topological charge is just the winding number of $\mathcal{W}$ as a map from $S^1$ to $\mathrm{O}(2)$, i.e.,
\begin{equation}\label{Eq:R21topo}
	\nu=\frac{1}{4\pi}\int_{S^1}d\phi~\mathrm{tr}J\mathcal W\partial_\phi\mathcal W^T \mod 2
\end{equation}
with the $\mathrm{SO}(2)$ generator $J=\mathrm{i}\sigma_2$. Particularly, for our model $\mathcal W(\phi)=[w_{+}(\phi)\sigma_1+w_{-}(\phi)\sigma_3]/|w(\phi)|$ leads to the nontrivial topological charge $\nu=1$.

{\color{blue}\textit{Summary}} 
In summary, we have established the theoretical foundation to study topological structures protected by CSNES. As bipartite lattices with CSNES widely exist, our work can be broadly applied to reveal novel topologies in crystalline systems with CSNES. Particularly, our theory has been applied to clarify the topological invariant of the PZSV model with both CSNES and PT symmetry. Models constructed here can be readily realized by various artificial crystals~\cite{Topo_ColdAtom_Review,Yu_Zhao_NSR,OzawaRMP2019,Peterson_2018nature,Serra_Garcia_2018nature,Liu2022nature,YangPRL2015,Xue2022NRM,LuNaturePhotonics2014,HuberNaturePhysics2016,Berry-dipole_Shuang}, and can pave the way for the further experimental investigation of novel topologies with CSNES. 

\bibliographystyle{apsrev4-1}
\bibliography{references}

\begin{thebibliography}{52}%
\makeatletter
\providecommand \@ifxundefined [1]{%
 \@ifx{#1\undefined}
}%
\providecommand \@ifnum [1]{%
 \ifnum #1\expandafter \@firstoftwo
 \else \expandafter \@secondoftwo
 \fi
}%
\providecommand \@ifx [1]{%
 \ifx #1\expandafter \@firstoftwo
 \else \expandafter \@secondoftwo
 \fi
}%
\providecommand \natexlab [1]{#1}%
\providecommand \enquote  [1]{``#1''}%
\providecommand \bibnamefont  [1]{#1}%
\providecommand \bibfnamefont [1]{#1}%
\providecommand \citenamefont [1]{#1}%
\providecommand \href@noop [0]{\@secondoftwo}%
\providecommand \href [0]{\begingroup \@sanitize@url \@href}%
\providecommand \@href[1]{\@@startlink{#1}\@@href}%
\providecommand \@@href[1]{\endgroup#1\@@endlink}%
\providecommand \@sanitize@url [0]{\catcode `\\12\catcode `\$12\catcode
  `\&12\catcode `\#12\catcode `\^12\catcode `\_12\catcode `\%12\relax}%
\providecommand \@@startlink[1]{}%
\providecommand \@@endlink[0]{}%
\providecommand \url  [0]{\begingroup\@sanitize@url \@url }%
\providecommand \@url [1]{\endgroup\@href {#1}{\urlprefix }}%
\providecommand \urlprefix  [0]{URL }%
\providecommand \Eprint [0]{\href }%
\providecommand \doibase [0]{http://dx.doi.org/}%
\providecommand \selectlanguage [0]{\@gobble}%
\providecommand \bibinfo  [0]{\@secondoftwo}%
\providecommand \bibfield  [0]{\@secondoftwo}%
\providecommand \translation [1]{[#1]}%
\providecommand \BibitemOpen [0]{}%
\providecommand \bibitemStop [0]{}%
\providecommand \bibitemNoStop [0]{.\EOS\space}%
\providecommand \EOS [0]{\spacefactor3000\relax}%
\providecommand \BibitemShut  [1]{\csname bibitem#1\endcsname}%
\let\auto@bib@innerbib\@empty
\bibitem [{\citenamefont {Regnault}\ \emph {et~al.}(2022)\citenamefont
  {Regnault}, \citenamefont {Xu}, \citenamefont {Li}, \citenamefont {Ma},
  \citenamefont {Jovanovic}, \citenamefont {Yazdani}, \citenamefont {Parkin},
  \citenamefont {Felser}, \citenamefont {Schoop}, \citenamefont {Ong} \emph
  {et~al.}}]{regnault2022catalogue}%
  \BibitemOpen
  \bibfield  {author} {\bibinfo {author} {\bibfnamefont {N.}~\bibnamefont
  {Regnault}}, \bibinfo {author} {\bibfnamefont {Y.}~\bibnamefont {Xu}},
  \bibinfo {author} {\bibfnamefont {M.-R.}\ \bibnamefont {Li}}, \bibinfo
  {author} {\bibfnamefont {D.-S.}\ \bibnamefont {Ma}}, \bibinfo {author}
  {\bibfnamefont {M.}~\bibnamefont {Jovanovic}}, \bibinfo {author}
  {\bibfnamefont {A.}~\bibnamefont {Yazdani}}, \bibinfo {author} {\bibfnamefont
  {S.~S.}\ \bibnamefont {Parkin}}, \bibinfo {author} {\bibfnamefont
  {C.}~\bibnamefont {Felser}}, \bibinfo {author} {\bibfnamefont {L.~M.}\
  \bibnamefont {Schoop}}, \bibinfo {author} {\bibfnamefont {N.~P.}\
  \bibnamefont {Ong}},  \emph {et~al.},\ }\href@noop {} {\bibfield  {journal}
  {\bibinfo  {journal} {Nature}\ }\textbf {\bibinfo {volume} {603}},\ \bibinfo
  {pages} {824} (\bibinfo {year} {2022})}\BibitemShut {NoStop}%
\bibitem [{\citenamefont {C{\u{a}}lug{\u{a}}ru}\ \emph
  {et~al.}(2022)\citenamefont {C{\u{a}}lug{\u{a}}ru}, \citenamefont {Chew},
  \citenamefont {Elcoro}, \citenamefont {Xu}, \citenamefont {Regnault},
  \citenamefont {Song},\ and\ \citenamefont
  {Bernevig}}]{cualuguaru2022general}%
  \BibitemOpen
  \bibfield  {author} {\bibinfo {author} {\bibfnamefont {D.}~\bibnamefont
  {C{\u{a}}lug{\u{a}}ru}}, \bibinfo {author} {\bibfnamefont {A.}~\bibnamefont
  {Chew}}, \bibinfo {author} {\bibfnamefont {L.}~\bibnamefont {Elcoro}},
  \bibinfo {author} {\bibfnamefont {Y.}~\bibnamefont {Xu}}, \bibinfo {author}
  {\bibfnamefont {N.}~\bibnamefont {Regnault}}, \bibinfo {author}
  {\bibfnamefont {Z.-D.}\ \bibnamefont {Song}}, \ and\ \bibinfo {author}
  {\bibfnamefont {B.~A.}\ \bibnamefont {Bernevig}},\ }\href@noop {} {\bibfield
  {journal} {\bibinfo  {journal} {Nature Physics}\ }\textbf {\bibinfo {volume}
  {18}},\ \bibinfo {pages} {185} (\bibinfo {year} {2022})}\BibitemShut
  {NoStop}%
\bibitem [{\citenamefont {Taie}\ \emph {et~al.}(2020)\citenamefont {Taie},
  \citenamefont {Ichinose}, \citenamefont {Ozawa},\ and\ \citenamefont
  {Takahashi}}]{taie2020spatial}%
  \BibitemOpen
  \bibfield  {author} {\bibinfo {author} {\bibfnamefont {S.}~\bibnamefont
  {Taie}}, \bibinfo {author} {\bibfnamefont {T.}~\bibnamefont {Ichinose}},
  \bibinfo {author} {\bibfnamefont {H.}~\bibnamefont {Ozawa}}, \ and\ \bibinfo
  {author} {\bibfnamefont {Y.}~\bibnamefont {Takahashi}},\ }\href@noop {}
  {\bibfield  {journal} {\bibinfo  {journal} {Nature Communications}\ }\textbf
  {\bibinfo {volume} {11}},\ \bibinfo {pages} {257} (\bibinfo {year}
  {2020})}\BibitemShut {NoStop}%
\bibitem [{\citenamefont {Slot}\ \emph {et~al.}(2017)\citenamefont {Slot},
  \citenamefont {Gardenier}, \citenamefont {Jacobse}, \citenamefont
  {Van~Miert}, \citenamefont {Kempkes}, \citenamefont {Zevenhuizen},
  \citenamefont {Smith}, \citenamefont {Vanmaekelbergh},\ and\ \citenamefont
  {Swart}}]{slot2017experimental}%
  \BibitemOpen
  \bibfield  {author} {\bibinfo {author} {\bibfnamefont {M.~R.}\ \bibnamefont
  {Slot}}, \bibinfo {author} {\bibfnamefont {T.~S.}\ \bibnamefont {Gardenier}},
  \bibinfo {author} {\bibfnamefont {P.~H.}\ \bibnamefont {Jacobse}}, \bibinfo
  {author} {\bibfnamefont {G.~C.}\ \bibnamefont {Van~Miert}}, \bibinfo {author}
  {\bibfnamefont {S.~N.}\ \bibnamefont {Kempkes}}, \bibinfo {author}
  {\bibfnamefont {S.~J.}\ \bibnamefont {Zevenhuizen}}, \bibinfo {author}
  {\bibfnamefont {C.~M.}\ \bibnamefont {Smith}}, \bibinfo {author}
  {\bibfnamefont {D.}~\bibnamefont {Vanmaekelbergh}}, \ and\ \bibinfo {author}
  {\bibfnamefont {I.}~\bibnamefont {Swart}},\ }\href@noop {} {\bibfield
  {journal} {\bibinfo  {journal} {Nature physics}\ }\textbf {\bibinfo {volume}
  {13}},\ \bibinfo {pages} {672} (\bibinfo {year} {2017})}\BibitemShut
  {NoStop}%
\bibitem [{\citenamefont {Vicencio}\ \emph {et~al.}(2015)\citenamefont
  {Vicencio}, \citenamefont {Cantillano}, \citenamefont {Morales-Inostroza},
  \citenamefont {Real}, \citenamefont {Mej\'{\i}a-Cort\'es}, \citenamefont
  {Weimann}, \citenamefont {Szameit},\ and\ \citenamefont
  {Molina}}]{Vicencio2015PRL}%
  \BibitemOpen
  \bibfield  {author} {\bibinfo {author} {\bibfnamefont {R.~A.}\ \bibnamefont
  {Vicencio}}, \bibinfo {author} {\bibfnamefont {C.}~\bibnamefont
  {Cantillano}}, \bibinfo {author} {\bibfnamefont {L.}~\bibnamefont
  {Morales-Inostroza}}, \bibinfo {author} {\bibfnamefont {B.}~\bibnamefont
  {Real}}, \bibinfo {author} {\bibfnamefont {C.}~\bibnamefont
  {Mej\'{\i}a-Cort\'es}}, \bibinfo {author} {\bibfnamefont {S.}~\bibnamefont
  {Weimann}}, \bibinfo {author} {\bibfnamefont {A.}~\bibnamefont {Szameit}}, \
  and\ \bibinfo {author} {\bibfnamefont {M.~I.}\ \bibnamefont {Molina}},\
  }\href@noop {} {\bibfield  {journal} {\bibinfo  {journal} {Phys. Rev. Lett.}\
  }\textbf {\bibinfo {volume} {114}},\ \bibinfo {pages} {245503} (\bibinfo
  {year} {2015})}\BibitemShut {NoStop}%
\bibitem [{\citenamefont {Mukherjee}\ \emph {et~al.}(2015)\citenamefont
  {Mukherjee}, \citenamefont {Spracklen}, \citenamefont {Choudhury},
  \citenamefont {Goldman}, \citenamefont {\"Ohberg}, \citenamefont
  {Andersson},\ and\ \citenamefont {Thomson}}]{Muk2015PRL}%
  \BibitemOpen
  \bibfield  {author} {\bibinfo {author} {\bibfnamefont {S.}~\bibnamefont
  {Mukherjee}}, \bibinfo {author} {\bibfnamefont {A.}~\bibnamefont
  {Spracklen}}, \bibinfo {author} {\bibfnamefont {D.}~\bibnamefont
  {Choudhury}}, \bibinfo {author} {\bibfnamefont {N.}~\bibnamefont {Goldman}},
  \bibinfo {author} {\bibfnamefont {P.}~\bibnamefont {\"Ohberg}}, \bibinfo
  {author} {\bibfnamefont {E.}~\bibnamefont {Andersson}}, \ and\ \bibinfo
  {author} {\bibfnamefont {R.~R.}\ \bibnamefont {Thomson}},\ }\href@noop {}
  {\bibfield  {journal} {\bibinfo  {journal} {Phys. Rev. Lett.}\ }\textbf
  {\bibinfo {volume} {114}},\ \bibinfo {pages} {245504} (\bibinfo {year}
  {2015})}\BibitemShut {NoStop}%
\bibitem [{\citenamefont {Xia}\ \emph {et~al.}(2018)\citenamefont {Xia},
  \citenamefont {Ramachandran}, \citenamefont {Xia}, \citenamefont {Li},
  \citenamefont {Liu}, \citenamefont {Tang}, \citenamefont {Hu}, \citenamefont
  {Song}, \citenamefont {Xu}, \citenamefont {Leykam}, \citenamefont {Flach},\
  and\ \citenamefont {Chen}}]{Xia2018PRL}%
  \BibitemOpen
  \bibfield  {author} {\bibinfo {author} {\bibfnamefont {S.}~\bibnamefont
  {Xia}}, \bibinfo {author} {\bibfnamefont {A.}~\bibnamefont {Ramachandran}},
  \bibinfo {author} {\bibfnamefont {S.}~\bibnamefont {Xia}}, \bibinfo {author}
  {\bibfnamefont {D.}~\bibnamefont {Li}}, \bibinfo {author} {\bibfnamefont
  {X.}~\bibnamefont {Liu}}, \bibinfo {author} {\bibfnamefont {L.}~\bibnamefont
  {Tang}}, \bibinfo {author} {\bibfnamefont {Y.}~\bibnamefont {Hu}}, \bibinfo
  {author} {\bibfnamefont {D.}~\bibnamefont {Song}}, \bibinfo {author}
  {\bibfnamefont {J.}~\bibnamefont {Xu}}, \bibinfo {author} {\bibfnamefont
  {D.}~\bibnamefont {Leykam}}, \bibinfo {author} {\bibfnamefont
  {S.}~\bibnamefont {Flach}}, \ and\ \bibinfo {author} {\bibfnamefont
  {Z.}~\bibnamefont {Chen}},\ }\href@noop {} {\bibfield  {journal} {\bibinfo
  {journal} {Phys. Rev. Lett.}\ }\textbf {\bibinfo {volume} {121}},\ \bibinfo
  {pages} {263902} (\bibinfo {year} {2018})}\BibitemShut {NoStop}%
\bibitem [{\citenamefont {Julku}\ \emph {et~al.}(2016)\citenamefont {Julku},
  \citenamefont {Peotta}, \citenamefont {Vanhala}, \citenamefont {Kim},\ and\
  \citenamefont {T\"orm\"a}}]{Julku2016PRL}%
  \BibitemOpen
  \bibfield  {author} {\bibinfo {author} {\bibfnamefont {A.}~\bibnamefont
  {Julku}}, \bibinfo {author} {\bibfnamefont {S.}~\bibnamefont {Peotta}},
  \bibinfo {author} {\bibfnamefont {T.~I.}\ \bibnamefont {Vanhala}}, \bibinfo
  {author} {\bibfnamefont {D.-H.}\ \bibnamefont {Kim}}, \ and\ \bibinfo
  {author} {\bibfnamefont {P.}~\bibnamefont {T\"orm\"a}},\ }\href@noop {}
  {\bibfield  {journal} {\bibinfo  {journal} {Phys. Rev. Lett.}\ }\textbf
  {\bibinfo {volume} {117}},\ \bibinfo {pages} {045303} (\bibinfo {year}
  {2016})}\BibitemShut {NoStop}%
\bibitem [{\citenamefont {Andrijauskas}\ \emph {et~al.}(2015)\citenamefont
  {Andrijauskas}, \citenamefont {Anisimovas}, \citenamefont {Ra\ifmmode
  \check{c}\else \v{c}\fi{}i\ifmmode~\bar{u}\else \={u}\fi{}nas}, \citenamefont
  {Mekys}, \citenamefont {Kudria\ifmmode~\check{s}\else \v{s}\fi{}ov},
  \citenamefont {Spielman},\ and\ \citenamefont {Juzeli\ifmmode~\bar{u}\else
  \={u}\fi{}nas}}]{Andrj2015PRA}%
  \BibitemOpen
  \bibfield  {author} {\bibinfo {author} {\bibfnamefont {T.}~\bibnamefont
  {Andrijauskas}}, \bibinfo {author} {\bibfnamefont {E.}~\bibnamefont
  {Anisimovas}}, \bibinfo {author} {\bibfnamefont {M.}~\bibnamefont {Ra\ifmmode
  \check{c}\else \v{c}\fi{}i\ifmmode~\bar{u}\else \={u}\fi{}nas}}, \bibinfo
  {author} {\bibfnamefont {A.}~\bibnamefont {Mekys}}, \bibinfo {author}
  {\bibfnamefont {V.}~\bibnamefont {Kudria\ifmmode~\check{s}\else
  \v{s}\fi{}ov}}, \bibinfo {author} {\bibfnamefont {I.~B.}\ \bibnamefont
  {Spielman}}, \ and\ \bibinfo {author} {\bibfnamefont {G.}~\bibnamefont
  {Juzeli\ifmmode~\bar{u}\else \={u}\fi{}nas}},\ }\href@noop {} {\bibfield
  {journal} {\bibinfo  {journal} {Phys. Rev. A}\ }\textbf {\bibinfo {volume}
  {92}},\ \bibinfo {pages} {033617} (\bibinfo {year} {2015})}\BibitemShut
  {NoStop}%
\bibitem [{\citenamefont {Mohanta}\ \emph {et~al.}(2023)\citenamefont
  {Mohanta}, \citenamefont {Soni}, \citenamefont {Okamoto},\ and\ \citenamefont
  {Dagotto}}]{mohanta2023majorana}%
  \BibitemOpen
  \bibfield  {author} {\bibinfo {author} {\bibfnamefont {N.}~\bibnamefont
  {Mohanta}}, \bibinfo {author} {\bibfnamefont {R.}~\bibnamefont {Soni}},
  \bibinfo {author} {\bibfnamefont {S.}~\bibnamefont {Okamoto}}, \ and\
  \bibinfo {author} {\bibfnamefont {E.}~\bibnamefont {Dagotto}},\ }\href@noop
  {} {\bibfield  {journal} {\bibinfo  {journal} {Communications Physics}\
  }\textbf {\bibinfo {volume} {6}},\ \bibinfo {pages} {240} (\bibinfo {year}
  {2023})}\BibitemShut {NoStop}%
\bibitem [{\citenamefont {Wang}\ and\ \citenamefont {Ran}(2011)}]{Wang2011PRB}%
  \BibitemOpen
  \bibfield  {author} {\bibinfo {author} {\bibfnamefont {F.}~\bibnamefont
  {Wang}}\ and\ \bibinfo {author} {\bibfnamefont {Y.}~\bibnamefont {Ran}},\
  }\href@noop {} {\bibfield  {journal} {\bibinfo  {journal} {Phys. Rev. B}\
  }\textbf {\bibinfo {volume} {84}},\ \bibinfo {pages} {241103} (\bibinfo
  {year} {2011})}\BibitemShut {NoStop}%
\bibitem [{\citenamefont {Po}\ \emph {et~al.}(2019)\citenamefont {Po},
  \citenamefont {Zou}, \citenamefont {Senthil},\ and\ \citenamefont
  {Vishwanath}}]{Vishwanath2019PRB}%
  \BibitemOpen
  \bibfield  {author} {\bibinfo {author} {\bibfnamefont {H.~C.}\ \bibnamefont
  {Po}}, \bibinfo {author} {\bibfnamefont {L.}~\bibnamefont {Zou}}, \bibinfo
  {author} {\bibfnamefont {T.}~\bibnamefont {Senthil}}, \ and\ \bibinfo
  {author} {\bibfnamefont {A.}~\bibnamefont {Vishwanath}},\ }\href@noop {}
  {\bibfield  {journal} {\bibinfo  {journal} {Phys. Rev. B}\ }\textbf {\bibinfo
  {volume} {99}},\ \bibinfo {pages} {195455} (\bibinfo {year}
  {2019})}\BibitemShut {NoStop}%
\bibitem [{\citenamefont {Graf}\ and\ \citenamefont
  {Pi\'echon}(2023)}]{Hopf_dipole}%
  \BibitemOpen
  \bibfield  {author} {\bibinfo {author} {\bibfnamefont {A.}~\bibnamefont
  {Graf}}\ and\ \bibinfo {author} {\bibfnamefont {F.}~\bibnamefont
  {Pi\'echon}},\ }\href@noop {} {\bibfield  {journal} {\bibinfo  {journal}
  {Phys. Rev. B}\ }\textbf {\bibinfo {volume} {108}},\ \bibinfo {pages}
  {115105} (\bibinfo {year} {2023})}\BibitemShut {NoStop}%
\bibitem [{\citenamefont {Mo}\ \emph {et~al.}(2024)\citenamefont {Mo},
  \citenamefont {Zheng}, \citenamefont {Lu}, \citenamefont {Huang},
  \citenamefont {Liu},\ and\ \citenamefont {Zhang}}]{Berry-dipole_Shuang}%
  \BibitemOpen
  \bibfield  {author} {\bibinfo {author} {\bibfnamefont {Q.}~\bibnamefont
  {Mo}}, \bibinfo {author} {\bibfnamefont {R.}~\bibnamefont {Zheng}}, \bibinfo
  {author} {\bibfnamefont {C.}~\bibnamefont {Lu}}, \bibinfo {author}
  {\bibfnamefont {X.}~\bibnamefont {Huang}}, \bibinfo {author} {\bibfnamefont
  {Z.}~\bibnamefont {Liu}}, \ and\ \bibinfo {author} {\bibfnamefont
  {S.}~\bibnamefont {Zhang}},\ }\href@noop {} {\bibfield  {journal} {\bibinfo
  {journal} {arXiv preprint arXiv:2405.07569}\ } (\bibinfo {year}
  {2024})}\BibitemShut {NoStop}%
\bibitem [{\citenamefont {Schnyder}\ \emph {et~al.}(2008)\citenamefont
  {Schnyder}, \citenamefont {Ryu}, \citenamefont {Furusaki},\ and\
  \citenamefont {Ludwig}}]{Schnyder2008PRB}%
  \BibitemOpen
  \bibfield  {author} {\bibinfo {author} {\bibfnamefont {A.~P.}\ \bibnamefont
  {Schnyder}}, \bibinfo {author} {\bibfnamefont {S.}~\bibnamefont {Ryu}},
  \bibinfo {author} {\bibfnamefont {A.}~\bibnamefont {Furusaki}}, \ and\
  \bibinfo {author} {\bibfnamefont {A.~W.~W.}\ \bibnamefont {Ludwig}},\
  }\href@noop {} {\bibfield  {journal} {\bibinfo  {journal} {Phys. Rev. B}\
  }\textbf {\bibinfo {volume} {78}},\ \bibinfo {pages} {195125} (\bibinfo
  {year} {2008})}\BibitemShut {NoStop}%
\bibitem [{\citenamefont {Atiyah}(1966)}]{Atiyah-KR}%
  \BibitemOpen
  \bibfield  {author} {\bibinfo {author} {\bibfnamefont {M.~F.}\ \bibnamefont
  {Atiyah}},\ }\href@noop {} {\bibfield  {journal} {\bibinfo  {journal} {The
  Quarterly Journal of Mathematics}\ }\textbf {\bibinfo {volume} {17}},\
  \bibinfo {pages} {367} (\bibinfo {year} {1966})}\BibitemShut {NoStop}%
\bibitem [{\citenamefont {Kitaev}(2010)}]{Kitaev2009AIP}%
  \BibitemOpen
  \bibfield  {author} {\bibinfo {author} {\bibfnamefont {A.}~\bibnamefont
  {Kitaev}},\ }\href@noop {} {\bibfield  {journal} {\bibinfo  {journal} {AIP
  Conference Proceedings}\ }\textbf {\bibinfo {volume} {1134}},\ \bibinfo
  {pages} {22} (\bibinfo {year} {2010})}\BibitemShut {NoStop}%
\bibitem [{\citenamefont {Altland}\ and\ \citenamefont
  {Zirnbauer}(1997)}]{AZ1997PRB}%
  \BibitemOpen
  \bibfield  {author} {\bibinfo {author} {\bibfnamefont {A.}~\bibnamefont
  {Altland}}\ and\ \bibinfo {author} {\bibfnamefont {M.~R.}\ \bibnamefont
  {Zirnbauer}},\ }\href@noop {} {\bibfield  {journal} {\bibinfo  {journal}
  {Phys. Rev. B}\ }\textbf {\bibinfo {volume} {55}},\ \bibinfo {pages} {1142}
  (\bibinfo {year} {1997})}\BibitemShut {NoStop}%
\bibitem [{\citenamefont {Zhao}\ and\ \citenamefont
  {Wang}(2013)}]{ZhaoYXWang13prl}%
  \BibitemOpen
  \bibfield  {author} {\bibinfo {author} {\bibfnamefont {Y.~X.}\ \bibnamefont
  {Zhao}}\ and\ \bibinfo {author} {\bibfnamefont {Z.~D.}\ \bibnamefont
  {Wang}},\ }\href@noop {} {\bibfield  {journal} {\bibinfo  {journal} {Phys.
  Rev. Lett.}\ }\textbf {\bibinfo {volume} {110}},\ \bibinfo {pages} {240404}
  (\bibinfo {year} {2013})}\BibitemShut {NoStop}%
\bibitem [{\citenamefont {Volovik}(2003)}]{Volovik:book}%
  \BibitemOpen
  \bibfield  {author} {\bibinfo {author} {\bibfnamefont {G.~E.}\ \bibnamefont
  {Volovik}},\ }\href@noop {} {\emph {\bibinfo {title} {Universe in a helium
  droplet}}}\ (\bibinfo  {publisher} {Oxford University Press, Oxford UK},\
  \bibinfo {year} {2003})\BibitemShut {NoStop}%
\bibitem [{\citenamefont {Qi}\ and\ \citenamefont
  {Zhang}(2011)}]{SCZHANGRMP2011}%
  \BibitemOpen
  \bibfield  {author} {\bibinfo {author} {\bibfnamefont {X.-L.}\ \bibnamefont
  {Qi}}\ and\ \bibinfo {author} {\bibfnamefont {S.-C.}\ \bibnamefont {Zhang}},\
  }\href@noop {} {\bibfield  {journal} {\bibinfo  {journal} {Rev. Mod. Phys.}\
  }\textbf {\bibinfo {volume} {83}},\ \bibinfo {pages} {1057} (\bibinfo {year}
  {2011})}\BibitemShut {NoStop}%
\bibitem [{\citenamefont {Shen}(2012)}]{Shun-Qing-Shen}%
  \BibitemOpen
  \bibfield  {author} {\bibinfo {author} {\bibfnamefont {S.-Q.}\ \bibnamefont
  {Shen}},\ }\href@noop {} {\emph {\bibinfo {title} {Topological Insulators:
  Dirac Equation in Condensed Matters}}}\ (\bibinfo  {publisher} {Springer},\
  \bibinfo {year} {2012})\ pp.\ \bibinfo {pages} {15--17}\BibitemShut {NoStop}%
\bibitem [{\citenamefont {Hasan}\ and\ \citenamefont
  {Kane}(2010)}]{KaneRMP2010}%
  \BibitemOpen
  \bibfield  {author} {\bibinfo {author} {\bibfnamefont {M.~Z.}\ \bibnamefont
  {Hasan}}\ and\ \bibinfo {author} {\bibfnamefont {C.~L.}\ \bibnamefont
  {Kane}},\ }\href@noop {} {\bibfield  {journal} {\bibinfo  {journal} {Rev.
  Mod. Phys.}\ }\textbf {\bibinfo {volume} {82}},\ \bibinfo {pages} {3045}
  (\bibinfo {year} {2010})}\BibitemShut {NoStop}%
\bibitem [{\citenamefont {Chiu}\ \emph {et~al.}(2016)\citenamefont {Chiu},
  \citenamefont {Teo}, \citenamefont {Schnyder},\ and\ \citenamefont
  {Ryu}}]{SchnyderRMP2016}%
  \BibitemOpen
  \bibfield  {author} {\bibinfo {author} {\bibfnamefont {C.-K.}\ \bibnamefont
  {Chiu}}, \bibinfo {author} {\bibfnamefont {J.~C.~Y.}\ \bibnamefont {Teo}},
  \bibinfo {author} {\bibfnamefont {A.~P.}\ \bibnamefont {Schnyder}}, \ and\
  \bibinfo {author} {\bibfnamefont {S.}~\bibnamefont {Ryu}},\ }\href@noop {}
  {\bibfield  {journal} {\bibinfo  {journal} {Rev. Mod. Phys.}\ }\textbf
  {\bibinfo {volume} {88}},\ \bibinfo {pages} {035005} (\bibinfo {year}
  {2016})}\BibitemShut {NoStop}%
\bibitem [{\citenamefont {Teo}\ and\ \citenamefont {Kane}(2010)}]{Teo2010PRB}%
  \BibitemOpen
  \bibfield  {author} {\bibinfo {author} {\bibfnamefont {J.~C.~Y.}\
  \bibnamefont {Teo}}\ and\ \bibinfo {author} {\bibfnamefont {C.~L.}\
  \bibnamefont {Kane}},\ }\href@noop {} {\bibfield  {journal} {\bibinfo
  {journal} {Phys. Rev. B}\ }\textbf {\bibinfo {volume} {82}},\ \bibinfo
  {pages} {115120} (\bibinfo {year} {2010})}\BibitemShut {NoStop}%
\bibitem [{\citenamefont {IM}(1977)}]{TopologySM}%
  \BibitemOpen
  \bibfield  {author} {\bibinfo {author} {\bibfnamefont {J.}~\bibnamefont
  {IM}},\ }\href@noop {} {\emph {\bibinfo {title} {The Topology of Stiefel
  Manifolds}}}\ (\bibinfo  {publisher} {Cambridge University Press},\ \bibinfo
  {year} {1977})\BibitemShut {NoStop}%
\bibitem [{Wik()}]{Wiki}%
  \BibitemOpen
  \href@noop {} {}\bibinfo {note} {See the wikipedia item:
  \url{https://en.wikipedia.org/wiki/Stiefel_manifold}}\BibitemShut {NoStop}%
\bibitem [{\citenamefont {Wang}\ \emph {et~al.}(2020)\citenamefont {Wang},
  \citenamefont {Dai}, \citenamefont {Shao}, \citenamefont {Yang},\ and\
  \citenamefont {Zhao}}]{WangPRL2020}%
  \BibitemOpen
  \bibfield  {author} {\bibinfo {author} {\bibfnamefont {K.}~\bibnamefont
  {Wang}}, \bibinfo {author} {\bibfnamefont {J.-X.}\ \bibnamefont {Dai}},
  \bibinfo {author} {\bibfnamefont {L.~B.}\ \bibnamefont {Shao}}, \bibinfo
  {author} {\bibfnamefont {S.~A.}\ \bibnamefont {Yang}}, \ and\ \bibinfo
  {author} {\bibfnamefont {Y.~X.}\ \bibnamefont {Zhao}},\ }\href@noop {}
  {\bibfield  {journal} {\bibinfo  {journal} {Phys. Rev. Lett.}\ }\textbf
  {\bibinfo {volume} {125}},\ \bibinfo {pages} {126403} (\bibinfo {year}
  {2020})}\BibitemShut {NoStop}%
\bibitem [{\citenamefont {Zhao}\ \emph {et~al.}(2016)\citenamefont {Zhao},
  \citenamefont {Schnyder},\ and\ \citenamefont {Wang}}]{YXZHAOPRL2016}%
  \BibitemOpen
  \bibfield  {author} {\bibinfo {author} {\bibfnamefont {Y.~X.}\ \bibnamefont
  {Zhao}}, \bibinfo {author} {\bibfnamefont {A.~P.}\ \bibnamefont {Schnyder}},
  \ and\ \bibinfo {author} {\bibfnamefont {Z.~D.}\ \bibnamefont {Wang}},\
  }\href@noop {} {\bibfield  {journal} {\bibinfo  {journal} {Phys. Rev. Lett.}\
  }\textbf {\bibinfo {volume} {116}},\ \bibinfo {pages} {156402} (\bibinfo
  {year} {2016})}\BibitemShut {NoStop}%
\bibitem [{\citenamefont {Zhao}\ and\ \citenamefont
  {Lu}(2017)}]{YXZHAOPRL2017}%
  \BibitemOpen
  \bibfield  {author} {\bibinfo {author} {\bibfnamefont {Y.~X.}\ \bibnamefont
  {Zhao}}\ and\ \bibinfo {author} {\bibfnamefont {Y.}~\bibnamefont {Lu}},\
  }\href@noop {} {\bibfield  {journal} {\bibinfo  {journal} {Phys. Rev. Lett.}\
  }\textbf {\bibinfo {volume} {118}},\ \bibinfo {pages} {056401} (\bibinfo
  {year} {2017})}\BibitemShut {NoStop}%
\bibitem [{\citenamefont {Ahn}\ \emph {et~al.}(2018)\citenamefont {Ahn},
  \citenamefont {Kim}, \citenamefont {Kim},\ and\ \citenamefont
  {Yang}}]{BJYPRL2018}%
  \BibitemOpen
  \bibfield  {author} {\bibinfo {author} {\bibfnamefont {J.}~\bibnamefont
  {Ahn}}, \bibinfo {author} {\bibfnamefont {D.}~\bibnamefont {Kim}}, \bibinfo
  {author} {\bibfnamefont {Y.}~\bibnamefont {Kim}}, \ and\ \bibinfo {author}
  {\bibfnamefont {B.-J.}\ \bibnamefont {Yang}},\ }\href@noop {} {\bibfield
  {journal} {\bibinfo  {journal} {Phys. Rev. Lett.}\ }\textbf {\bibinfo
  {volume} {121}},\ \bibinfo {pages} {106403} (\bibinfo {year}
  {2018})}\BibitemShut {NoStop}%
\bibitem [{\citenamefont {Ahn}\ \emph {et~al.}(2019)\citenamefont {Ahn},
  \citenamefont {Park},\ and\ \citenamefont {Yang}}]{BJYPRX2019}%
  \BibitemOpen
  \bibfield  {author} {\bibinfo {author} {\bibfnamefont {J.}~\bibnamefont
  {Ahn}}, \bibinfo {author} {\bibfnamefont {S.}~\bibnamefont {Park}}, \ and\
  \bibinfo {author} {\bibfnamefont {B.-J.}\ \bibnamefont {Yang}},\ }\href@noop
  {} {\bibfield  {journal} {\bibinfo  {journal} {Phys. Rev. X}\ }\textbf
  {\bibinfo {volume} {9}},\ \bibinfo {pages} {021013} (\bibinfo {year}
  {2019})}\BibitemShut {NoStop}%
\bibitem [{\citenamefont {Bzdusek}\ and\ \citenamefont
  {Sigrist}(2017)}]{Tomas2017PRB}%
  \BibitemOpen
  \bibfield  {author} {\bibinfo {author} {\bibfnamefont {T.}~\bibnamefont
  {Bzdusek}}\ and\ \bibinfo {author} {\bibfnamefont {M.}~\bibnamefont
  {Sigrist}},\ }\href@noop {} {\bibfield  {journal} {\bibinfo  {journal} {Phys.
  Rev. B}\ }\textbf {\bibinfo {volume} {96}},\ \bibinfo {pages} {155105}
  (\bibinfo {year} {2017})}\BibitemShut {NoStop}%
\bibitem [{\citenamefont {Xue}\ \emph {et~al.}(2023)\citenamefont {Xue},
  \citenamefont {Chen}, \citenamefont {Cheng}, \citenamefont {Dai},
  \citenamefont {Long}, \citenamefont {Zhao},\ and\ \citenamefont
  {Zhang}}]{xue2023stiefel}%
  \BibitemOpen
  \bibfield  {author} {\bibinfo {author} {\bibfnamefont {H.}~\bibnamefont
  {Xue}}, \bibinfo {author} {\bibfnamefont {Z.}~\bibnamefont {Chen}}, \bibinfo
  {author} {\bibfnamefont {Z.}~\bibnamefont {Cheng}}, \bibinfo {author}
  {\bibfnamefont {J.}~\bibnamefont {Dai}}, \bibinfo {author} {\bibfnamefont
  {Y.}~\bibnamefont {Long}}, \bibinfo {author} {\bibfnamefont {Y.}~\bibnamefont
  {Zhao}}, \ and\ \bibinfo {author} {\bibfnamefont {B.}~\bibnamefont {Zhang}},\
  }\href@noop {} {\bibfield  {journal} {\bibinfo  {journal} {Nature
  Communications}\ }\textbf {\bibinfo {volume} {14}} (\bibinfo {year}
  {2023})}\BibitemShut {NoStop}%
\bibitem [{\citenamefont {Dai}\ \emph {et~al.}(2021)\citenamefont {Dai},
  \citenamefont {Wang}, \citenamefont {Yang},\ and\ \citenamefont
  {Zhao}}]{DAI2021PRB}%
  \BibitemOpen
  \bibfield  {author} {\bibinfo {author} {\bibfnamefont {J.-X.}\ \bibnamefont
  {Dai}}, \bibinfo {author} {\bibfnamefont {K.}~\bibnamefont {Wang}}, \bibinfo
  {author} {\bibfnamefont {S.~A.}\ \bibnamefont {Yang}}, \ and\ \bibinfo
  {author} {\bibfnamefont {Y.~X.}\ \bibnamefont {Zhao}},\ }\href@noop {}
  {\bibfield  {journal} {\bibinfo  {journal} {Phys. Rev. B}\ }\textbf {\bibinfo
  {volume} {104}},\ \bibinfo {pages} {165142} (\bibinfo {year}
  {2021})}\BibitemShut {NoStop}%
\bibitem [{\citenamefont {Sheng}\ \emph {et~al.}(2019)\citenamefont {Sheng},
  \citenamefont {Chen}, \citenamefont {Liu}, \citenamefont {Chen},
  \citenamefont {Yu}, \citenamefont {Zhao},\ and\ \citenamefont
  {Yang}}]{ShengXLPRL2019}%
  \BibitemOpen
  \bibfield  {author} {\bibinfo {author} {\bibfnamefont {X.-L.}\ \bibnamefont
  {Sheng}}, \bibinfo {author} {\bibfnamefont {C.}~\bibnamefont {Chen}},
  \bibinfo {author} {\bibfnamefont {H.}~\bibnamefont {Liu}}, \bibinfo {author}
  {\bibfnamefont {Z.}~\bibnamefont {Chen}}, \bibinfo {author} {\bibfnamefont
  {Z.-M.}\ \bibnamefont {Yu}}, \bibinfo {author} {\bibfnamefont {Y.~X.}\
  \bibnamefont {Zhao}}, \ and\ \bibinfo {author} {\bibfnamefont {S.~A.}\
  \bibnamefont {Yang}},\ }\href@noop {} {\bibfield  {journal} {\bibinfo
  {journal} {Phys. Rev. Lett.}\ }\textbf {\bibinfo {volume} {123}},\ \bibinfo
  {pages} {256402} (\bibinfo {year} {2019})}\BibitemShut {NoStop}%
\bibitem [{\citenamefont {Dyson}(1962)}]{dyson}%
  \BibitemOpen
  \bibfield  {author} {\bibinfo {author} {\bibfnamefont {F.~J.}\ \bibnamefont
  {Dyson}},\ }\href@noop {} {\bibfield  {journal} {\bibinfo  {journal} {Journal
  of Mathematical Physics}\ }\textbf {\bibinfo {volume} {3}},\ \bibinfo {pages}
  {1199} (\bibinfo {year} {1962})}\BibitemShut {NoStop}%
\bibitem [{Co-()}]{Co-dim}%
  \BibitemOpen
  \href@noop {} {}\bibinfo {note} {The codimension is defined as the
  dimensionality of the sphere $S^d$ enclosing the gapless modes from their
  transverse dimensions in momentum space.}\BibitemShut {Stop}%
\bibitem [{Sup()}]{Supp}%
  \BibitemOpen
  \href@noop {} {}\bibinfo {note} {See the Supplemental Material for
  topological classifications, topological invariants, matrix representation of
  spin-1 operators, a 1D insulator model with CSNES and $\mathcal P\mathcal T$
  symmetry, and ten-band PZSV lattice model with CSNES and $\mathcal P\mathcal
  T$ symmetry}\BibitemShut {NoStop}%
\bibitem [{\citenamefont {Zhu}\ \emph {et~al.}(2017)\citenamefont {Zhu},
  \citenamefont {Zhang}, \citenamefont {Yan}, \citenamefont {Xing},\ and\
  \citenamefont {Zhu}}]{ZHUYANQING2017PRA}%
  \BibitemOpen
  \bibfield  {author} {\bibinfo {author} {\bibfnamefont {Y.-Q.}\ \bibnamefont
  {Zhu}}, \bibinfo {author} {\bibfnamefont {D.-W.}\ \bibnamefont {Zhang}},
  \bibinfo {author} {\bibfnamefont {H.}~\bibnamefont {Yan}}, \bibinfo {author}
  {\bibfnamefont {D.-Y.}\ \bibnamefont {Xing}}, \ and\ \bibinfo {author}
  {\bibfnamefont {S.-L.}\ \bibnamefont {Zhu}},\ }\href@noop {} {\bibfield
  {journal} {\bibinfo  {journal} {Phys. Rev. A}\ }\textbf {\bibinfo {volume}
  {96}},\ \bibinfo {pages} {033634} (\bibinfo {year} {2017})}\BibitemShut
  {NoStop}%
\bibitem [{\citenamefont {Tan}\ \emph {et~al.}(2018)\citenamefont {Tan},
  \citenamefont {Zhang}, \citenamefont {Liu}, \citenamefont {Xue},
  \citenamefont {Yu}, \citenamefont {Zhu}, \citenamefont {Yan}, \citenamefont
  {Zhu},\ and\ \citenamefont {Yu}}]{TAN2018PRL}%
  \BibitemOpen
  \bibfield  {author} {\bibinfo {author} {\bibfnamefont {X.}~\bibnamefont
  {Tan}}, \bibinfo {author} {\bibfnamefont {D.-W.}\ \bibnamefont {Zhang}},
  \bibinfo {author} {\bibfnamefont {Q.}~\bibnamefont {Liu}}, \bibinfo {author}
  {\bibfnamefont {G.}~\bibnamefont {Xue}}, \bibinfo {author} {\bibfnamefont
  {H.-F.}\ \bibnamefont {Yu}}, \bibinfo {author} {\bibfnamefont {Y.-Q.}\
  \bibnamefont {Zhu}}, \bibinfo {author} {\bibfnamefont {H.}~\bibnamefont
  {Yan}}, \bibinfo {author} {\bibfnamefont {S.-L.}\ \bibnamefont {Zhu}}, \ and\
  \bibinfo {author} {\bibfnamefont {Y.}~\bibnamefont {Yu}},\ }\href@noop {}
  {\bibfield  {journal} {\bibinfo  {journal} {Phys. Rev. Lett.}\ }\textbf
  {\bibinfo {volume} {120}},\ \bibinfo {pages} {130503} (\bibinfo {year}
  {2018})}\BibitemShut {NoStop}%
\bibitem [{\citenamefont {Stone}(2016)}]{Stone2016IJMPB}%
  \BibitemOpen
  \bibfield  {author} {\bibinfo {author} {\bibfnamefont {M.}~\bibnamefont
  {Stone}},\ }\href@noop {} {\bibfield  {journal} {\bibinfo  {journal}
  {International Journal of Modern Physics B}\ }\textbf {\bibinfo {volume}
  {30}},\ \bibinfo {pages} {1550249} (\bibinfo {year} {2016})}\BibitemShut
  {NoStop}%
\bibitem [{\citenamefont {Zhang}\ \emph {et~al.}(2018)\citenamefont {Zhang},
  \citenamefont {Zhu}, \citenamefont {Zhao}, \citenamefont {Yan},\ and\
  \citenamefont {Zhu}}]{Topo_ColdAtom_Review}%
  \BibitemOpen
  \bibfield  {author} {\bibinfo {author} {\bibfnamefont {D.-W.}\ \bibnamefont
  {Zhang}}, \bibinfo {author} {\bibfnamefont {Y.-Q.}\ \bibnamefont {Zhu}},
  \bibinfo {author} {\bibfnamefont {Y.~X.}\ \bibnamefont {Zhao}}, \bibinfo
  {author} {\bibfnamefont {H.}~\bibnamefont {Yan}}, \ and\ \bibinfo {author}
  {\bibfnamefont {S.-L.}\ \bibnamefont {Zhu}},\ }\href@noop {} {\bibfield
  {journal} {\bibinfo  {journal} {Advances in Physics}\ }\textbf {\bibinfo
  {volume} {67}},\ \bibinfo {pages} {253} (\bibinfo {year} {2018})}\BibitemShut
  {NoStop}%
\bibitem [{\citenamefont {Yu}\ \emph {et~al.}(2020)\citenamefont {Yu},
  \citenamefont {Zhao},\ and\ \citenamefont {Schnyder}}]{Yu_Zhao_NSR}%
  \BibitemOpen
  \bibfield  {author} {\bibinfo {author} {\bibfnamefont {R.}~\bibnamefont
  {Yu}}, \bibinfo {author} {\bibfnamefont {Y.~X.}\ \bibnamefont {Zhao}}, \ and\
  \bibinfo {author} {\bibfnamefont {A.~P.}\ \bibnamefont {Schnyder}},\
  }\href@noop {} {\bibfield  {journal} {\bibinfo  {journal} {National Science
  Review}\ } (\bibinfo {year} {2020})},\ \bibinfo {note} {nwaa065}\BibitemShut
  {NoStop}%
\bibitem [{\citenamefont {Ozawa}\ \emph {et~al.}(2019)\citenamefont {Ozawa},
  \citenamefont {Price}, \citenamefont {Amo}, \citenamefont {Goldman},
  \citenamefont {Hafezi}, \citenamefont {Lu}, \citenamefont {Rechtsman},
  \citenamefont {Schuster}, \citenamefont {Simon}, \citenamefont {Zilberberg},\
  and\ \citenamefont {Carusotto}}]{OzawaRMP2019}%
  \BibitemOpen
  \bibfield  {author} {\bibinfo {author} {\bibfnamefont {T.}~\bibnamefont
  {Ozawa}}, \bibinfo {author} {\bibfnamefont {H.~M.}\ \bibnamefont {Price}},
  \bibinfo {author} {\bibfnamefont {A.}~\bibnamefont {Amo}}, \bibinfo {author}
  {\bibfnamefont {N.}~\bibnamefont {Goldman}}, \bibinfo {author} {\bibfnamefont
  {M.}~\bibnamefont {Hafezi}}, \bibinfo {author} {\bibfnamefont
  {L.}~\bibnamefont {Lu}}, \bibinfo {author} {\bibfnamefont {M.~C.}\
  \bibnamefont {Rechtsman}}, \bibinfo {author} {\bibfnamefont {D.}~\bibnamefont
  {Schuster}}, \bibinfo {author} {\bibfnamefont {J.}~\bibnamefont {Simon}},
  \bibinfo {author} {\bibfnamefont {O.}~\bibnamefont {Zilberberg}}, \ and\
  \bibinfo {author} {\bibfnamefont {I.}~\bibnamefont {Carusotto}},\ }\href@noop
  {} {\bibfield  {journal} {\bibinfo  {journal} {Rev. Mod. Phys.}\ }\textbf
  {\bibinfo {volume} {91}},\ \bibinfo {pages} {015006} (\bibinfo {year}
  {2019})}\BibitemShut {NoStop}%
\bibitem [{\citenamefont {Peterson}\ \emph {et~al.}(2018)\citenamefont
  {Peterson}, \citenamefont {Benalcazar}, \citenamefont {Hughes},\ and\
  \citenamefont {Bahl}}]{Peterson_2018nature}%
  \BibitemOpen
  \bibfield  {author} {\bibinfo {author} {\bibfnamefont {C.~W.}\ \bibnamefont
  {Peterson}}, \bibinfo {author} {\bibfnamefont {W.~A.}\ \bibnamefont
  {Benalcazar}}, \bibinfo {author} {\bibfnamefont {T.~L.}\ \bibnamefont
  {Hughes}}, \ and\ \bibinfo {author} {\bibfnamefont {G.}~\bibnamefont
  {Bahl}},\ }\href@noop {} {\bibfield  {journal} {\bibinfo  {journal} {Nature}\
  }\textbf {\bibinfo {volume} {555}},\ \bibinfo {pages} {346} (\bibinfo {year}
  {2018})}\BibitemShut {NoStop}%
\bibitem [{\citenamefont {Serra-Garcia}\ \emph {et~al.}(2018)\citenamefont
  {Serra-Garcia}, \citenamefont {Peri}, \citenamefont {Süsstrunk},
  \citenamefont {Bilal}, \citenamefont {Larsen}, \citenamefont {Villanueva},\
  and\ \citenamefont {Huber}}]{Serra_Garcia_2018nature}%
  \BibitemOpen
  \bibfield  {author} {\bibinfo {author} {\bibfnamefont {M.}~\bibnamefont
  {Serra-Garcia}}, \bibinfo {author} {\bibfnamefont {V.}~\bibnamefont {Peri}},
  \bibinfo {author} {\bibfnamefont {R.}~\bibnamefont {Süsstrunk}}, \bibinfo
  {author} {\bibfnamefont {O.~R.}\ \bibnamefont {Bilal}}, \bibinfo {author}
  {\bibfnamefont {T.}~\bibnamefont {Larsen}}, \bibinfo {author} {\bibfnamefont
  {L.~G.}\ \bibnamefont {Villanueva}}, \ and\ \bibinfo {author} {\bibfnamefont
  {S.~D.}\ \bibnamefont {Huber}},\ }\href@noop {} {\bibfield  {journal}
  {\bibinfo  {journal} {Nature}\ }\textbf {\bibinfo {volume} {555}},\ \bibinfo
  {pages} {342} (\bibinfo {year} {2018})}\BibitemShut {NoStop}%
\bibitem [{\citenamefont {Liu}\ \emph {et~al.}(2022)\citenamefont {Liu},
  \citenamefont {Gao}, \citenamefont {Wang}, \citenamefont {Xi}, \citenamefont
  {Hu}, \citenamefont {Wang}, \citenamefont {Liu}, \citenamefont {Lin},
  \citenamefont {Deng}, \citenamefont {Yang}, \citenamefont {Zhou},
  \citenamefont {Yang}, \citenamefont {Chong},\ and\ \citenamefont
  {Zhang}}]{Liu2022nature}%
  \BibitemOpen
  \bibfield  {author} {\bibinfo {author} {\bibfnamefont {G.~G.}\ \bibnamefont
  {Liu}}, \bibinfo {author} {\bibfnamefont {Z.}~\bibnamefont {Gao}}, \bibinfo
  {author} {\bibfnamefont {Q.}~\bibnamefont {Wang}}, \bibinfo {author}
  {\bibfnamefont {X.}~\bibnamefont {Xi}}, \bibinfo {author} {\bibfnamefont
  {Y.~H.}\ \bibnamefont {Hu}}, \bibinfo {author} {\bibfnamefont
  {M.}~\bibnamefont {Wang}}, \bibinfo {author} {\bibfnamefont {C.}~\bibnamefont
  {Liu}}, \bibinfo {author} {\bibfnamefont {X.}~\bibnamefont {Lin}}, \bibinfo
  {author} {\bibfnamefont {L.}~\bibnamefont {Deng}}, \bibinfo {author}
  {\bibfnamefont {S.~A.}\ \bibnamefont {Yang}}, \bibinfo {author}
  {\bibfnamefont {P.}~\bibnamefont {Zhou}}, \bibinfo {author} {\bibfnamefont
  {Y.}~\bibnamefont {Yang}}, \bibinfo {author} {\bibfnamefont {Y.}~\bibnamefont
  {Chong}}, \ and\ \bibinfo {author} {\bibfnamefont {B.}~\bibnamefont
  {Zhang}},\ }\href@noop {} {\bibfield  {journal} {\bibinfo  {journal}
  {Nature}\ }\textbf {\bibinfo {volume} {609}},\ \bibinfo {pages} {925}
  (\bibinfo {year} {2022})}\BibitemShut {NoStop}%
\bibitem [{\citenamefont {Yang}\ \emph {et~al.}(2015)\citenamefont {Yang},
  \citenamefont {Gao}, \citenamefont {Shi}, \citenamefont {Lin}, \citenamefont
  {Gao}, \citenamefont {Chong},\ and\ \citenamefont {Zhang}}]{YangPRL2015}%
  \BibitemOpen
  \bibfield  {author} {\bibinfo {author} {\bibfnamefont {Z.}~\bibnamefont
  {Yang}}, \bibinfo {author} {\bibfnamefont {F.}~\bibnamefont {Gao}}, \bibinfo
  {author} {\bibfnamefont {X.}~\bibnamefont {Shi}}, \bibinfo {author}
  {\bibfnamefont {X.}~\bibnamefont {Lin}}, \bibinfo {author} {\bibfnamefont
  {Z.}~\bibnamefont {Gao}}, \bibinfo {author} {\bibfnamefont {Y.}~\bibnamefont
  {Chong}}, \ and\ \bibinfo {author} {\bibfnamefont {B.}~\bibnamefont
  {Zhang}},\ }\href@noop {} {\bibfield  {journal} {\bibinfo  {journal} {Phys.
  Rev. Lett.}\ }\textbf {\bibinfo {volume} {114}},\ \bibinfo {pages} {114301}
  (\bibinfo {year} {2015})}\BibitemShut {NoStop}%
\bibitem [{\citenamefont {Xue}\ \emph {et~al.}(2022)\citenamefont {Xue},
  \citenamefont {Yang},\ and\ \citenamefont {Zhang}}]{Xue2022NRM}%
  \BibitemOpen
  \bibfield  {author} {\bibinfo {author} {\bibfnamefont {H.}~\bibnamefont
  {Xue}}, \bibinfo {author} {\bibfnamefont {Y.}~\bibnamefont {Yang}}, \ and\
  \bibinfo {author} {\bibfnamefont {B.}~\bibnamefont {Zhang}},\ }\href@noop {}
  {\bibfield  {journal} {\bibinfo  {journal} {Nature Reviews Materials}\
  }\textbf {\bibinfo {volume} {7}},\ \bibinfo {pages} {974} (\bibinfo {year}
  {2022})}\BibitemShut {NoStop}%
\bibitem [{\citenamefont {Lu}\ \emph {et~al.}(2014)\citenamefont {Lu},
  \citenamefont {Joannopoulos},\ and\ \citenamefont
  {Solja{\v{c}}i{\'c}}}]{LuNaturePhotonics2014}%
  \BibitemOpen
  \bibfield  {author} {\bibinfo {author} {\bibfnamefont {L.}~\bibnamefont
  {Lu}}, \bibinfo {author} {\bibfnamefont {J.~D.}\ \bibnamefont
  {Joannopoulos}}, \ and\ \bibinfo {author} {\bibfnamefont {M.}~\bibnamefont
  {Solja{\v{c}}i{\'c}}},\ }\href@noop {} {\bibfield  {journal} {\bibinfo
  {journal} {Nature photonics}\ }\textbf {\bibinfo {volume} {8}},\ \bibinfo
  {pages} {821} (\bibinfo {year} {2014})}\BibitemShut {NoStop}%
\bibitem [{\citenamefont {Huber}(2016)}]{HuberNaturePhysics2016}%
  \BibitemOpen
  \bibfield  {author} {\bibinfo {author} {\bibfnamefont {S.~D.}\ \bibnamefont
  {Huber}},\ }\href@noop {} {\bibfield  {journal} {\bibinfo  {journal} {Nature
  Physics}\ }\textbf {\bibinfo {volume} {12}},\ \bibinfo {pages} {621}
  (\bibinfo {year} {2016})}\BibitemShut {NoStop}%
\end{thebibliography}%

\end{document}


\title{The Supplemental Materials for ``Topological classification of chiral symmetry with non-equal sublattices''}

\author{J. X. Dai}
\affiliation{Department of Physics and HKU-UCAS Joint Institute for Theoretical
	and Computational Physics at Hong Kong, The University of Hong Kong,
	Pokfulam Road, Hong Kong, China}
\affiliation{HK Institute of Quantum Science \& Technology, The University of Hong Kong,
	Pokfulam Road, Hong Kong, China}

\author{Y. X. Zhao}
\email[]{yuxinphy@hku.hk}
\affiliation{Department of Physics and HKU-UCAS Joint Institute for Theoretical
	and Computational Physics at Hong Kong, The University of Hong Kong,
	Pokfulam Road, Hong Kong, China}
\affiliation{HK Institute of Quantum Science \& Technology, The University of Hong Kong,
	Pokfulam Road, Hong Kong, China}
\maketitle
%

\setcounter{page}{1}
\makeatletter
\renewcommand\theequation{S\arabic{equation}}
\renewcommand\thefigure{S\arabic{figure}}
\setcounter{equation}{0}

\section{Topological classifications}
\subsection{The classifying space: Stiefel manifolds}
In our manuscript, we discuss bipartite systems that maintain chiral symmetry across non-equal sublattices, referred to as chiral symmetry with non-equal sublattices (CSNES). The CSNE operator can be represented by
\begin{equation}
	\mathcal{S}=\begin{bmatrix}
		1_M & 0\\
		0 & -1_N
	\end{bmatrix}.
\end{equation} 
Here, we assume $M<N$ without loss of generality for subsequent discussions. The Hamiltonian $\H(\k)$ and the operator $\S$ satisfy an anti-commutation relation, which restricts $\H(\k)$ to an anti-diagonal form

\begin{equation}\label{eq:anti-diagonal}
	\H(\k)=\begin{bmatrix}
		&\Q(\k)\\\Q^\dagger(\k)&
	\end{bmatrix}.
\end{equation}
In this equation, $\Q(\k)$ is an $M \times N$ matrix and $\Q^\dagger(\k)$ is its conjugate transpose, an $N \times M$ matrix.

The matrix $\Q(\k)$ can be factorized via its singular value decomposition (SVD) as:
\begin{equation}\label{eq:SVD}
	\Q(\k)=\mathcal V(\k)\mathcal{D}(\k)\U^\dagger(\k).
\end{equation}
Here, $\mathcal{V}$ and $\mathcal{U}$ are unitary matrices of rank $M$ and $N$, respectively. The matrix $\mathcal{D}$ is of size $M\times N$ and structured as $\mathcal{D}=[\Lambda,\mathbf{0}]$, where $\Lambda$ is a diagonal $M\times M$ matrix and $\mathbf{0}$ is a zero $M\times(N-M)$ matrix. The diagonal entries of $\Lambda$ are the singular values $\lambda_i$ of $\Q$, where $\lambda_i\ge 0$. Utilizing the SVD, the Hamiltonian $\H(\k)$ can be expressed in decomposed form,
\begin{equation}\label{Eq:Uni-trans}
	\H(\bm{k})=\begin{bmatrix}
		\mathcal V(\k)&\\&\U(\k)
	\end{bmatrix}\begin{bmatrix}
		&\mathcal{D}(\bm{k})\\\mathcal{D}^\dagger(\bm{k})&
	\end{bmatrix}\begin{bmatrix}
		\mathcal V^\dagger(\k)&\\&\U^\dagger(\k)
	\end{bmatrix}.
\end{equation}
This decomposition implies that each singular value $\lambda_i$ contributes a pair of eigenvalues $\pm\lambda_i$ to $\H(\k)$. 

To analyze the topological properties of $\H(\k)$, we transform it into its flattened version $\widetilde\H(\k)$ by continuously deforming the eigenvalues $\lambda_i(\k)$ to $1$. In this flattened representation, the eigenvalues of $\H(\k)$ are confined to $0$ and $\pm 1$. We denote the flattened version of any operator $\O$ as $\widetilde{\O}$. Specially, for $\H(\k)$, we have $\widetilde{\Lambda}=1_{M}$ and $\widetilde{\mathcal{D}}=(1_M, 0_{M\times(N-M)})$. Consequently, we can express $\widetilde{\Q}$ as:
\begin{equation}\label{eq:SimplifiedSVD1}
	\widetilde{\Q}=[1_M, \mathbf{0}]\begin{bmatrix}
		\mathcal V &\\ & 1_{N-M}
	\end{bmatrix}\U^\dagger.
\end{equation}
Therefore, any $\Q$ can be represented in its flattened form:
\begin{equation}\label{eq:canonical_form}
	\widetilde\Q=[1_M, \mathbf{0}]\mathcal{W},
\end{equation}
where $\mathcal{W}$ can be any $N\times N$ unitary matrix, i.e., $\W\in \mathrm{U}(N)$. It is crucial to recognize that $\mathcal{W}$ is not uniquely determined by $\widetilde{\Q}$, as $\widetilde{\Q}$ is invariant under the transformation
\begin{equation}\label{Eq:Gauge}
	\mathcal W\mapsto \begin{bmatrix}
		1_M &\\ & \mathcal{G}
	\end{bmatrix}\mathcal W,
\end{equation}
where $\mathcal{G}$ is any unitary matrix with rank $N-M$, i.e., $\mathcal{G}\in \mathrm U(N-M)$. The inverse is also true. If $\mathcal{W}$ and $\mathcal{W}'$ correspond to the same $\widetilde{\Q}$, there exists a unitary matrix $\mathcal{G}\in \mathrm{U}(N-M)$ with $\mathcal{W}'=\mathrm{diag}[1_M,\mathcal{G}]\mathcal{W}$. Thus, we obtain the representation of the complex Stiefel manifold
\begin{equation}\label{Eq:Complexmanifold}
	V_M(\mathbb C^N)=\mathrm{U}(N)/\mathrm{U}(N-M).
\end{equation}

If the spacetime inversion symmetry $\P\T$ with $(PT)^2=1$ [$(PT)^2=-1$] is added on the bipartite system, $\W$ in Eq.~\eqref{eq:canonical_form} and $\G$ in Eq.~\eqref{Eq:Gauge} would be real orthogonal (symplectic) matrices. And the classifying space would become the real (quaternionic) Stiefel manifold $V_M(\mathbb{R}^N)$ [$V_M(\mathbb{H}^N)$], with
\begin{equation}
	V_M(\mathbb R^N)=\mathrm{O}(N)/\mathrm{O}(N-M),~V_M(\mathbb H^N)=\mathrm{Sp}(N)/\mathrm{Sp}(N-M).
\end{equation}
The topological classifications of the Stiefel manifolds can be calculated as their homotopy groups. In the following, we would show the calculation details of the homotopy groups of the Stiefel manifolds.

\subsection{Homotopy groups of Stiefel manifolds}
 It is observed that all Stiefel manifolds take the form of a quotient space $X/A$. To ascertain the homotopy groups of $X/A$, we employ the long exact sequence arising from the short exact sequence of the form
\begin{equation}
	0\rightarrow A\rightarrow X\rightarrow X/A\rightarrow 0.
\end{equation}
This sequence induces a long exact sequence for homotopy groups:
\begin{equation}\label{eq:longexact}
	\cdots\rightarrow\pi_n(A)\rightarrow\pi_n(X)\rightarrow\pi_n(X/A)\rightarrow\pi_{n-1}(A)\rightarrow\pi_{n-1}(X)\rightarrow\pi_{n-1}(X/A)\rightarrow\cdots.
\end{equation}
Utilizing the sequence \eqref{eq:longexact}, we deduce the topological classifications of the Stiefopy manifolds, as illustrated in Table.~\ref{Table:Realtable}. Next, we shall present the calculation details for the homotopy groups of the complex, real, and quaternionic Stiefel manifolds, which are essential for the topological classification.
\begin{table}[t]
	\begin{tabular}{c|c|c| c c c}
		$(\P\T)^2$ & $V_M(\mathbb{K}^N)$ & $N-M$ & $d=1$ & $d=2$  & $d=3$\\
		\hline
		\hline
		0  &  ~$V_M(\mathbb C^{N})$~  & 0  & $\Z$&0&$\Z$\\
		& &1 & 0&0&$\Z$\\
		\hline
		1  & $V_M(\mathbb R^{N})$   &0 & $\Z_2$&0&$\Z$\\
		& &1  & $\Z_2$&0&$\Z$\\
		& &2 & 0&$2\Z$&$\Z$\\
		& &3  & 0&0&$\Z_2$\\
		\hline
		-1 & $V_M(\mathbb H^{N})$  &0 & $0$&0&$\Z$
	\end{tabular}
	\caption{Topological classification table of the Stiefel manifolds. The first column specifies the three cases of $\P\T$ symmetry, which determines the classifying space $V_M[\mathbb{K}^N]$. Here, $(\P\T)^2=0$ indicates the absence of $\P\T$ symmetry. For each classifying space, $N-M$ is fixed at various values while $N$ and $M$ are presumed to be sufficiently large. The corresponding homotopy groups $\pi_d(V_M(\mathbb{K}^N))$ are presented for $d=1,2,3$. Note that $\pi_d[V_M(\mathbb{K}^N)]=0$ if $N-M$ is greater than the exhibited range.}\label{Table:Realtable}
\end{table}
\subsubsection{Complex Stiefel manifolds}
As shown in Table~\ref{Table:Realtable}, we establish that $\pi_{1}[V_M(\mathbb{C}^N)]=0$, $\pi_{2}[V_M(\mathbb{C}^N)]=0$, $\pi_{3}[V_{M}(\mathbb{C}^N)]=0$ with $M\leq N-2$, and $\pi_{3}[V_{N-1}(\mathbb{C}^N)]\cong\mathbb{Z}$. Now, let us provide the derivation details for these consequences.

\begin{itemize}
	\item For $\pi_{1}[V_M(\mathbb{C}^N)]$, the exact sequence is structured as follows:
	\begin{equation}\label{eq:cmes1}
		\cdots\rightarrow\underset{\mathbb{Z}}{\pi_1[\mathrm{U}(N-M)]}\xrightarrow{i_1}\underset{\mathbb{Z}}{\pi_1[\mathrm{U}(N)]}\xrightarrow{i_2}\pi_1[V_M(\mathbb{C}^N)]\xrightarrow{i_3}\underset{0}{\pi_0[\mathrm{U}(N-M)]}\rightarrow\cdots.
	\end{equation}
	Using the fact that the image of $i_j$ equals the kernel of $i_{j+1}$ for exact sequences, we can conclude that 
	\begin{equation}\label{eq:cmki1}
		\text{ker}(i_3)=\text{im}(i_2)\cong\mathbb{Z}/\text{ker}(i_2)=\mathbb{Z}/\text{im}(i_1)=0,~\text{im}(i_3)=0.
	\end{equation} 
	Here, we utilize the fact that $i_1$ is an isomorphism from $\mathbb{Z}$ to $\mathbb{Z}$, which implies $\mathbb{Z}/\text{im}(i_1)=0$. Equations~\eqref{eq:cmes1} and \eqref{eq:cmki1} indicate that $i_3$ is an isomorphism from $\pi_1[V_M(\mathbb{C}^N)]$ to the trivial group $0$. Therefore, we obtain that
	\begin{equation}
		\pi_1[V_M(\mathbb{C}^N)]=0.
	\end{equation}
	
	\item For $\pi_{2}[V_M(\mathbb{C}^N)]$, the exact sequence is structured as follows:
	\begin{equation}\label{eq:cmes2}
		\cdots\rightarrow\pi_2[\mathrm U(N-M)]\xrightarrow{i_1}\pi_2[\mathrm U(N)]\xrightarrow{i_2}\pi_2[V_M(\mathbb C^N)]\xrightarrow{i_3}\pi_1[\mathrm U(N-M)]\xrightarrow{i_4}\pi_1[\mathrm U(N)]\rightarrow\cdots.
	\end{equation}
	In this condition, we know that 
	\begin{equation}\label{eq:cmki2}
		\begin{split}
			\text{ker}(i_3)=\text{im}(i_2)=0,~\text{im}(i_3)=\text{ker}(i_4)=0.
		\end{split}
	\end{equation}
	Here, we utilize the fact that $i_4$ is an isomorphism from $\Z$ to $\Z$, which indicates that $\text{ker}(i_4)=0$. Then, we obtain that
	\begin{equation}
		\pi_2[V_M(\mathbb C^N)]=0.
	\end{equation}
	
	\item For $\pi_{3}[V_M(\mathbb{C}^N)]$ with $M\leq N-2$, the exact sequence is structured as follows:  
	\begin{equation}\label{eq:cmes3}
		\cdots\rightarrow\pi_3[\mathrm U(N-M)]\xrightarrow{i_1}\pi_3[\mathrm U(N)]\xrightarrow{i_2}\pi_3[V_M(\mathbb C^N)]\xrightarrow{i_3}\pi_2[\mathrm U(N-M)]\rightarrow\cdots.
	\end{equation}
	In this condition, we know that 
	\begin{equation}\label{eq:cmki3}
		\begin{split}
			\text{ker}(i_3)=\text{im}(i_2)\cong\Z/\text{ker}(i_2)=\Z/\text{im}(i_1)=0,~\text{im}(i_3)=0.
		\end{split}
	\end{equation}
	Here, we utilize the fact that $i_1$ is an isomorphism from $\Z$ to $\Z$, which implies $\Z/\text{im}(i_1)=0$. Then, we obtain that
	\begin{equation}
		\pi_3[V_M(\mathbb C^N)]=0~\text{with}~M\leq N-2.
	\end{equation}                                                                                          
	
	\item For $\pi_{3}[V_{N-1}(\mathbb C^N)]$, the exact sequence is structured as follows:
	\begin{equation}\label{eq:cmes4}
		\cdots\rightarrow\pi_3[\mathrm U(1)]\xrightarrow{i_1}\pi_3[\mathrm U(N)]\xrightarrow{i_2}\pi_3[V_{N-1}(\mathbb C^N)]\xrightarrow{i_3}\pi_2[\mathrm U(1)]\rightarrow\cdots.
	\end{equation}
	In this condition, we know that 
	\begin{equation}\label{eq:cmki4}
		\begin{split}
			\text{ker}(i_3)=\text{im}(i_2)\cong\Z/\text{ker}(i_2)=\Z/\text{im}(i_1)\cong\Z,~\text{im}(i_3)=0.
		\end{split}
	\end{equation}
	Since $\pi_3[V_{N-1}(\mathbb C^N)]/\text{ker}(i_3)\cong\text{im}(i_3)$, we obtain that
	\begin{equation}
		\pi_3[V_{N-1}(\mathbb C^N)]\cong\Z.
	\end{equation}
\end{itemize}

\subsubsection{Real Stiefel manifolds}
As shown in Table.~\ref{Table:Realtable}, we have $\pi_{1}[V_{N-1}(\mathbb R^N)]\cong\Z_2$, $\pi_{1}[V_{M}(\mathbb R^N)]=0$ with $M\leq N-2$, $\pi_{2}[V_{N-1}(\mathbb R^N)]=0$, $\pi_{2}[V_{N-2}(\mathbb R^N)]\cong2\Z$, $\pi_{2}[V_{M}(\mathbb R^N)]=0$ with $M\leq N-3$, $\pi_{3}[V_{M}(\mathbb R^N)]\cong\Z$ with $M\geq N-2$, $\pi_{3}[V_{N-3}(\mathbb R^N)]\cong\Z_2$, and $\pi_{3}[V_{M}(\mathbb R^N)]=0$ with $M\leq N-4$. Now, let us provide the derivation details for these results.
\begin{itemize}
	\item For $\pi_{1}[V_{N-1}(\mathbb R^N)]$, the exact sequence can be represented as follows:
	\begin{equation}\label{eq:rmes1}
		\cdots\rightarrow\pi_1[\mathrm O(1)]\xrightarrow{i_1}\pi_1[\mathrm O(N)]\xrightarrow{i_2}\pi_1[V_{N-1}(\mathbb R^N)]\xrightarrow{i_3}\pi_0[\mathrm O(1)]\xrightarrow{i_4}\pi_0[\mathrm O(N)]\rightarrow\cdots.
	\end{equation}
	From this sequence, we can deduce the following relationships:
	\begin{equation}\label{eq:rmki1}
		\text{ker}(i_3)=\text{im}(i_2)\cong\Z_2/\text{ker}(i_2)=\Z_2/\text{im}(i_1)\cong\Z_2,~\text{im}(i_3)=\text{ker}(i_4)=0.
	\end{equation}
	We utilize the fact that $i_4$ is an isomorphism from $\Z_2$ to $\Z_2$, which implies $\text{ker}(i_4)=0$. Since $\pi_1[V_{N-1}(\mathbb R^N)]/\text{ker}(i_3)\cong\text{im}(i_3)$, we can conclude that:
	\begin{equation}
		\pi_1[V_{N-1}(\mathbb R^N)]\cong\Z_2.
	\end{equation}
	
	\item For $\pi_{1}[V_{M}(\mathbb R^N)]$ with $M\leq N-2$, the exact sequence can be represented as follows:
	\begin{equation}\label{eq:rmes2}
		\cdots\rightarrow\pi_1[\mathrm O(N-M)]\xrightarrow{i_1}\pi_1[\mathrm O(N)]\xrightarrow{i_2}\pi_1[V_{N-1}(\mathbb R^N)]\xrightarrow{i_3}\pi_0[\mathrm O(N-M)]\xrightarrow{i_4}\pi_0[\mathrm O(N)]\rightarrow\cdots.
	\end{equation}
	From this sequence, we can deduce the following relationships:
	\begin{equation}\label{eq:rmki2}
		\text{ker}(i_3)=\text{im}(i_2)\cong\Z_2/\text{ker}(i_2)=\Z_2/\text{im}(i_1)=0,~\text{im}(i_3)=\text{ker}(i_4)=0.
	\end{equation}
	We utilize the fact that $i_1$ is an isomorphism from $\Z_2$ to $\Z_2$, which implies $\Z_2/\text{im}(i_1)=0$. Equation~\eqref{eq:rmki2} indicates that:
	\begin{equation}
		\pi_1[V_{M}(\mathbb R^N)]=0~\text{for}~M\leq N-2.
	\end{equation}
	
	\item For $\pi_{2}[V_{N-1}(\mathbb R^N)]$, the exact sequence can be represented as follows:
	\begin{equation}\label{eq:rmes3}
		\cdots\rightarrow\pi_2[\mathrm O(N)]\xrightarrow{i_1}\pi_2[V_{N-1}(\mathbb R^N)]\xrightarrow{i_2}\pi_1[\mathrm O(1)]\rightarrow\cdots.
	\end{equation}
	From this sequence, we can deduce the following relationships:
	\begin{equation}\label{eq:rmki3}
		\text{ker}(i_2)=\text{im}(i_1)=0,~\text{im}(i_2)=0.
	\end{equation}
	Equation~\eqref{eq:rmki3} indicates that:
	\begin{equation}
		\pi_2[V_{N-1}(\mathbb R^N)]=0.
	\end{equation}
	
	\item For $\pi_{2}[V_{N-2}(\mathbb R^N)]$, the exact sequence can be represented as follows:
	\begin{equation}\label{eq:rmes4}
		\cdots\rightarrow\pi_2[\mathrm O(N)]\xrightarrow{i_1}\pi_2[V_{N-2}(\mathbb R^N)]\xrightarrow{i_2}\pi_1[\mathrm O(2)]\xrightarrow{i_3}\pi_1[\mathrm O(N)]\rightarrow\cdots.
	\end{equation}
	From this sequence, we can deduce the following relationships:
	\begin{equation}\label{eq:rmki4}
		\text{ker}(i_2)=\text{im}(i_1)=0,~\text{im}(i_2)=\text{ker}(i_3)\cong2\Z.
	\end{equation}
	We utilize the fact that $i_3$ maps the even/odd numbers of $\Z$ to $0$/$1$ of $\Z_2$, which leads to $\text{ker}(i_3)\cong2\Z$. Since $\pi_2[V_{N-2}(\mathbb R^N)]/\text{ker}(i_2)\cong\text{im}(i_2)$, we can conclude that:
	\begin{equation}
		\pi_2[V_{N-2}(\mathbb R^N)]\cong2\Z.
	\end{equation}

	\item For $\pi_{2}[V_{M}(\mathbb R^N)]$ with $M\leq N-3$, the exact sequence is structured as follows:   
	\begin{equation}\label{eq:rmes5}
		\cdots\rightarrow\pi_2[\mathrm O(N)]\xrightarrow{i_1}\pi_2[V_{M}(\mathbb R^N)]\xrightarrow{i_2}\pi_1[\mathrm O(N-M)]\xrightarrow{i_3}\pi_1[\mathrm O(N)]\rightarrow\cdots.
	\end{equation}
	Then, we know that 
	\begin{equation}\label{eq:rmki5}
		\text{ker}(i_2)=\text{im}(i_1)=0,~\text{im}(i_2)=\text{ker}(i_3)=0.
	\end{equation} 
	Here, we utilize the fact that $i_3$ is an isomorphism from $\Z_2$ to $\Z_2$, which implies that $\text{ker}(i_3)=0$. Equation~\eqref{eq:rmki5} indicates that 
	\begin{equation}
		\pi_2[V_{M}(\mathbb R^N)]=0~\text{for}~M\leq N-3.
	\end{equation}
	
	\item For $\pi_{3}[V_{M}(\mathbb R^N)]$ with $M\geq N-2$, the exact sequence is structured as follows: 
	\begin{equation}\label{eq:rmes6}
		\cdots\rightarrow\pi_3[\mathrm O(N-M)]\xrightarrow{i_1}\pi_3[\mathrm O(N)]\xrightarrow{i_2}\pi_3[V_{M}(\mathbb R^N)]\xrightarrow{i_3}\pi_2[\mathrm O(N-M)]\xrightarrow{i_4}\pi_2[\mathrm O(N)]\rightarrow\cdots.
	\end{equation}
	Then, we know that 
	\begin{equation}\label{eq:rmki6}
		\text{ker}(i_3)=\text{im}(i_2)=\Z/\text{ker}(i_2)=\Z/\text{im}(i_1)=\Z,~\text{im}(i_3)=0.
	\end{equation} 
	Since $\pi_3[V_{M}(\mathbb R^N)]/\text{ker}(i_3)\cong\text{im}(i_3)$, we obtain that 
	\begin{equation}
		\pi_3[V_{M}(\mathbb R^N)]\cong\Z~\text{for}~M\geq N-2.
	\end{equation}
	
	\item For $\pi_{3}[V_{N-3}(\mathbb R^N)]$, the exact sequence is structured as follows: 
	\begin{equation}\label{eq:rmes7}
		\cdots\rightarrow\pi_3[\mathrm O(3)]\xrightarrow{i_1}\pi_3[\mathrm O(N)]\xrightarrow{i_2}\pi_3[V_{N-3}(\mathbb R^N)]\xrightarrow{i_3}\pi_2[\mathrm O(3)]\xrightarrow{i_4}\pi_2[\mathrm O(N)]\rightarrow\cdots.
	\end{equation}
	Then, we know that 
	\begin{equation}\label{eq:rmki7}
		\text{ker}(i_3)=\text{im}(i_2)=\Z/\text{ker}(i_2)=\Z/\text{im}(i_1)\cong\Z_2,~\text{im}(i_3)=0.
	\end{equation} 
	Here, we utilize the fact that $i_1$ maps $2\Z$ to the even numbers of $\Z$, which implies that $\Z/\text{im}(i_1)\cong\Z_2$. Since $\pi_3[V_{N-3}(\mathbb R^N)]/\text{ker}(i_3)\cong\text{im}(i_3)$, we obtain that 
	\begin{equation}
		\pi_3[V_{N-3}(\mathbb R^N)]\cong\Z_2.
	\end{equation}

	\item For $\pi_{3}[V_{M}(\mathbb R^N)]$ with $M\leq N-4$, the exact sequence is structured as follows:  
	\begin{equation}\label{eq:rmes8}
		\cdots\rightarrow\pi_3[\mathrm O(N-M)]\xrightarrow{i_1}\pi_3[\mathrm O(N)]\xrightarrow{i_2}\pi_3[V_{M}(\mathbb R^N)]\xrightarrow{i_3}\pi_2[\mathrm O(N-M)]\xrightarrow{i_4}\pi_2[\mathrm O(N)]\rightarrow\cdots.
	\end{equation}
	Then, we know that 
	\begin{equation}\label{eq:rmki8}
		\text{ker}(i_3)=\text{im}(i_2)=\Z/\text{ker}(i_2)=\Z/\text{im}(i_1)=0,~\text{im}(i_3)=0.
	\end{equation} 
	Here, we utilize the fact that $i_1$ is an isomorphism from $\Z$ to $\Z$, which implies that $\Z/\text{im}(i_1)=0$. Then, we obtain that 
	\begin{equation}
		\pi_3[V_{M}(\mathbb R^N)]=0~\text{for}~M\leq N-4.
	\end{equation}
\end{itemize}

\subsubsection{Quaternionic Stiefel manifolds}
As shown in Table~\ref{Table:Realtable}, we have $\pi_{1}[V_{M}(\mathbb{H}^N)]=0$, $\pi_{2}[V_{M}(\mathbb{H}^N)]=0$, and $\pi_{3}[V_{M}(\mathbb{H}^N)]=0$. We will now present the derivation details that lead to these conclusions.
\begin{itemize}
	\item For $\pi_{1}[V_{M}(\mathbb{H}^N)]$, the exact sequence is structured as follows: 
	\begin{equation}\label{eq:qmes1}
		\cdots\rightarrow\underset{0}{\pi_1[\mathrm {Sp}(N)]}\xrightarrow{i_1}\pi_1[V_{M}(\mathbb{H}^N)]\xrightarrow{i_2}\underset{0}{\pi_0[\mathrm {Sp}(N-M)]}\rightarrow\cdots.
	\end{equation}
	From this, it follows that 
	\begin{equation}\label{eq:qmki1}
		\text{ker}(i_2)=\text{im}(i_1)=0,~\text{im}(i_2)=0.
	\end{equation} 
	Hence, it can be deduced that
	\begin{equation}
		\pi_1[V_{M}(\mathbb{H}^N)]=0.
	\end{equation}
	
	\item For $\pi_{2}[V_{M}(\mathbb H^N)]$, the exact sequence is structured as follows: 
	\begin{equation}\label{eq:qmes2}
		\cdots\rightarrow\underset{0}{\pi_2[\mathrm {Sp}(N)]}\xrightarrow{i_1}\pi_2[V_{N-1}(\mathbb H^N)]\xrightarrow{i_2}\underset{0}{\pi_1
			[\mathrm {Sp}(N-M)]}\rightarrow\cdots.
	\end{equation}
	Based on this sequence, it can be established that 
	\begin{equation}\label{eq:qmki2}
		\text{ker}(i_2)=\text{im}(i_1)=0,~\text{im}(i_2)=0.
	\end{equation} 
	The above equation indicates that 
	\begin{equation}
		\pi_2[V_{M}(\mathbb{H}^N)]=0.
	\end{equation}
	
	\item  For $\pi_{3}[V_{M}(\mathbb H^N)]$, the exact sequence is structured as follows: 
	\begin{equation}\label{eq:qmes3}
		\cdots\rightarrow\underset{\Z}{\pi_3[\mathrm {Sp}(N-M)]}\xrightarrow{i_1}\underset{\Z}{\pi_3[\mathrm {Sp}(N)]}\xrightarrow{i_2}\pi_3[V_{N-1}(\mathbb{H}^N)]\xrightarrow{i_3}\underset{0}{\pi_2
			[\mathrm {Sp}(N-M)]}\rightarrow\cdots.
	\end{equation}
	Based on this sequence, it can be inferred that 
	\begin{equation}\label{eq:qmki3}
		\text{ker}(i_3)=\text{im}(i_2)=\Z/\text{ker}(i_2)=\Z/\text{im}(i_1)=0,~\text{im}(i_3)=0.
	\end{equation} 
	Here, we utilize the fact that $i_1$ is an isomorphism from $\Z$ to $\Z$, which implies that $\Z/\text{im}(i_1)=0$. Equation~\eqref{eq:qmki3} indicates that 
	\begin{equation}
		\pi_3[V_{M}(\mathbb{H}^N)]=0.
	\end{equation}
\end{itemize}

\section{Topological invariants}
We will now give the methods for calculating the topological invariants for nontrivial topological classifications.

\subsection{$\pi_3[V_{N-1}(\mathbb{C}^N)]\cong\Z$}
Recalling Eq.~\eqref{eq:cmes4} and the fact that $\pi_3[V_{N-1}(\mathbb{C}^N)]\cong\Z$, we can obtain the following exact sequence:
\begin{equation}\label{topo_inva1}
	\underset{0}{\pi_3[\mathrm U(1)]}\xrightarrow{i_1}\underset{\Z}{\pi_3[\mathrm U(N)]}\xrightarrow{i_2}\underset{\Z}{\pi_3[V_{N-1}(\mathbb{C}^N)]}\xrightarrow{i_3}\underset{0}{\pi_2[\mathrm U(1)]}.
\end{equation}
As $i_2$ is an isomorphism from $\Z$ to $\Z$, the topological invariant $\mathcal N$ originates from $\pi_3[\mathrm U(N)]\cong\Z$. Since $\pi_2[\mathrm U(1)]=0$, a globally well-defined $\W(\k)\in\mathrm U(N)$, given by Eq.~\eqref{eq:canonical_form}, exists throughout the Brillouin zone. Therefore, $\mathcal N$ is simply the 3D winding number of the globally well-defined $\mathcal W(\k)$, which is expressed as follows:
\begin{equation}\label{21winding}
	\mathcal N=\frac{1}{24\pi^2}\int_{\mathrm{BZ}}\mathrm dk^3~\epsilon^{ijk}\mathrm{tr}~\W\partial_i\W^\dagger\W\partial_j\W^\dagger\W\partial_k\W^\dagger\in\Z.
\end{equation}

\subsection{$\pi_1[V_{N-1}(\mathbb R^N)]\cong\Z_2$}
Recalling Eq.~\eqref{eq:rmes1} and the fact that $\pi_1[V_{N-1}(\mathbb{R}^N)]\cong\Z_2$, we can obtain the following exact sequence:
\begin{equation}\label{topo_inva2}
	\underset{0}{\pi_1[\mathrm O(1)]}\xrightarrow{i_1}\underset{\Z_2}{\pi_1[\mathrm O(N)]}\xrightarrow{i_2}\underset{\Z_2}{\pi_1[V_{N-1}(\mathbb{R}^N)]}\xrightarrow{i_3}\underset{\Z_2}{\pi_0[\mathrm O(1)]}\xrightarrow{i_4}\underset{\Z_2}{\pi_0[\mathrm O(N)]}\rightarrow\cdots.
\end{equation}
As $i_3$ maps $\pi_1[V_{N-1}(\mathbb{R}^N)]$ to the trivial element of $\pi_1[\mathrm O(2)]\cong\Z_2$, $i_2$ is an isomorphism from $\Z_2$ to $\Z_2$. Consequently, the topological invariant $\mathcal N$ originates from $\pi_1[\mathrm O(N)]\cong\Z_2$. Since $i_3$ maps $\pi_1[V_{N-1}(\mathbb{R}^N)]\cong\Z_2$ to the trivial element of $\pi_0[\mathrm O(1)]\cong\Z_2$, there exists a globally well-defined $\W(k)\in\mathrm O(N)$, given by Eq.~\eqref{eq:canonical_form}, throughout the Brillouin zone. Therefore, $\mathcal N$ is simply the homotopy invariant of the globally well-defined $\mathcal W(k)$. When $N=2$, $\mathcal N$ can be computed as
\begin{equation}\label{R21winding}
	\mathcal N=\frac{1}{4\pi}\int_{T^1}\mathrm dk~\mathrm{tr}~\mathrm{i}\sigma_2\W\partial_k\W^T\mod 2.
\end{equation}
When $N>2$, $\mathcal N$ is equal to the parity of the number of times that the phases of the eigenvalues of $\W(k)$ cross $\pi$.

\subsection{$\pi_2[V_{N-2}(\mathbb R^N)]\cong2\Z$}
\begin{figure}[t]
	\includegraphics[width=2.5in]{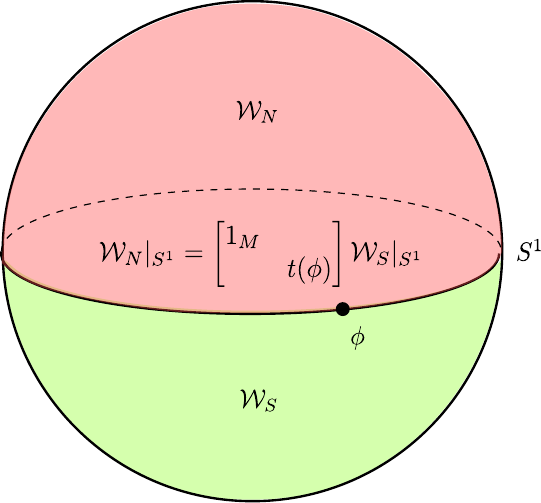}
	\caption{A sphere $S^2$ is divided into two hemisphere $D_N^2$ and $D_S^2$.}
	\label{S2}
\end{figure}
Recalling Eq.~\eqref{eq:rmes4} and the fact that $\pi_2[V_{N-2}(\mathbb{R}^N)]\cong2\Z$, we can obtain the following exact sequence:
\begin{equation}\label{topo_inva3}
	\underset{0}{\pi_2[\mathrm O(N)]}\xrightarrow{i_1}\underset{2\Z}{\pi_2[V_{N-2}(\mathbb{R}^N)]}\xrightarrow{i_2}\underset{\Z}{\pi_1[\mathrm O(2)]}\xrightarrow{i_3}\underset{\Z_2}{\pi_1[\mathrm O(N)]}\xrightarrow{i_4}\underset{0}{\pi_1[V_{N-2}(\mathbb{R}^N)]}.
\end{equation}
It is evident that $i_2$ maps $\pi_2[V_{N-2}(\mathbb{R}^N)]\cong2\Z$ to the even numbers of $\pi_1[\mathrm O(2)]\cong\Z$. Thus, the topological invariant $\mathcal N$ actually originates from $\pi_1[\mathrm O(2)]\cong\Z$.

Consider the 2D sphere $S^2$ in Fig.~\ref{S2}, which is divided into northern and southern hemispheres $D_{\mathrm{N/S}}^2$ with their intersection at $S^1$. We label $\W(\k)$, given by Eq.~\eqref{eq:canonical_form}, as $\W_\mathrm{N/S}$ on the hemisphere $D_\mathrm{N/S}^2$. The transition function $t(\phi)\in\mathrm O(2)$ glues $\W_\mathrm{N/S}$ along the equator $S^1$, as follows:
\begin{equation}\label{eq:transition}
	\W_\mathrm N|_{S^1}=\begin{bmatrix}
		1_{N-2}&\\& t(\phi)
	\end{bmatrix}\W_\mathrm{S}|_{S^1}.
\end{equation}
Then, $\pi_1[\mathrm O(2)]\cong\Z$ imposes obstructions to the globally well-defined $\mathcal W(\k)$. Thus, we can calculate the topological invariant $\mathcal N$ as the winding number of the transition function $t(\phi)$, that is,
\begin{equation}\label{R31winding}
	\mathcal N=\frac{1}{4\pi}\int_{T^1}\mathrm d\phi~\mathrm{tr}~\mathrm{i}\sigma_2t\partial_\phi t^T\in2\Z.
\end{equation}
It is noteworthy that the winding number of $t(\phi)$ must be an even number. We can also compute the phase factors $\omega_{1/2}$ of the eigenvalues $\chi_{1,2}$ of $t(\phi)$. $\mathcal N$ is equal to the number of times that $\omega_i$ crosses $\pi$.


\subsection{$\pi_3[V_{N-1}(\mathbb R^N)]\cong\Z$ and $\pi_3[V_{N-2}(\mathbb R^N)]\cong\Z$}
Recalling Eq.~\eqref{eq:rmes6} and the fact that $\pi_3[V_{M}(\mathbb R^N)]\cong\Z$ with $M\leq N-2$, we obtain the following sequence:
\begin{equation}\label{topo_inva4}
	\underset{0}{\pi_3[\mathrm O(N-M)]}\xrightarrow{i_1}\underset{\Z}{\pi_3[\mathrm O(N)]}\xrightarrow{i_2}\underset{\Z}{\pi_3[V_{M}(\mathbb R^N)]}\xrightarrow{i_3}\underset{0}{\pi_2[\mathrm O(N-M)]}.
\end{equation}
It is evident that $i_2$ is an isomorphism from $\Z$ to $\Z$. Therefore, the topological invariant $\mathcal N$ originates from $\pi_3[\mathrm O(N)]\cong\Z$. Furthermore, since $\pi_2[\mathrm O(N-M)]=0$, a globally well-defined $\W(\k)\in\mathrm O(N)$, which is defined in Equation~\eqref{eq:canonical_form}, always exists in the entire Brillouin zone. Consequently, $\mathcal N$ is the 3D winding number of the globally well-defined $\mathcal W(\k)$, which is given by:
\begin{equation}\label{R31winding3D}
	\mathcal N=\frac{1}{48\pi^2}\int_{\mathrm{BZ}}\mathrm dk^3~\epsilon^{ijk}\mathrm{tr}~\W\partial_i\W^T\W\partial_j\W^T\W\partial_k\W^T\in\Z.
\end{equation}

\subsection{$\pi_3[V_{N-3}(\mathbb R^N)]\cong\Z_2$}
Recalling Equation~\eqref{eq:rmes7} and the fact that $\pi_3[V_{N-3}(\mathbb R^N)]\cong\Z_2$, we obtain the following exact sequence:
\begin{equation}\label{topo_inva5}
	\underset{0}{\pi_4[V_{N-3}(\mathbb R^N)]}\xrightarrow{i_1}\underset{2\Z}{\pi_3[\mathrm O(3)]}\xrightarrow{i_2}\underset{\Z}{\pi_3[\mathrm O(N)]}\xrightarrow{i_3}\underset{\Z_2}{\pi_3[V_{N-3}(\mathbb R^N)]}\xrightarrow{i_4}\underset{0}{\pi_2[\mathrm O(3)]}.
\end{equation}
In the above equation, since $i_2$ maps $\pi_3[\mathrm{O}(3)]\cong2\Z$ to the even numbers of $\pi_3[\mathrm{O}(N)]\cong\Z$, $i_3$ maps even/odd numbers of $\Z$ to the trivial/nontrivial element of $\pi_3[\mathrm{O}(N)]\cong\Z$. Additionally, since $\pi_2[\mathrm O(3)]=0$, a globally well-defined $\W(\k)\in\mathrm O(N)$, as given in Equation~\eqref{eq:canonical_form}, always exists in the entire Brillouin zone. Consequently, the topological invariant $\mathcal N$ is the parity of the 3D winding number of the globally well-defined $\mathcal W(\k)\in\mathrm O(N)$, expressed as:
\begin{equation}\label{R31winding3D2}
	\mathcal N=\frac{1}{48\pi^2}\int_{\mathrm{BZ}}\mathrm dk^3~\epsilon^{ijk}\mathrm{tr}~\W\partial_i\W^T\W\partial_j\W^T\W\partial_k\W^T\mod 2.
\end{equation}

The reduction of the invariant from $\Z$ to $\Z_2$ is due to the requirement that a topological invariant for $\W$ should remain unchanged under any gauge transformation \eqref{Eq:Gauge}. Through straightforward derivations, it can be shown that:
\begin{equation}
	\mathcal N[\mathcal W']=\mathcal N[\W]+\mathcal N[\mathcal G]
\end{equation}
where $\mathcal W'=\mathrm{diag}[1_M,\mathcal G]\mathcal W$ and $\mathcal G\in\mathrm{O}(3)$. When substituting a topologically nontrivial $\mathcal G$ into the formula \eqref{R31winding3D2}, an even integer $\nu[\mathcal G]\in 2\Z$ is obtained, which is consistent with $\pi_3[\mathrm O(N)]\cong 2\Z$. Therefore, only the parity of the integer is meaningful here, justifying the $\Z_2$ nature of the topological invariant \eqref{R31winding3D2}.

\section{Matrix representation of spin-$1$ operators}
The matrices $S_{x,y,z}$ in the manuscript are given by
\begin{equation}
	S_x=\frac{1}{\sqrt{2}}\begin{bmatrix}
		0 & 1 & 1\\1 & 0 & 0\\1 & 0 & 0
	\end{bmatrix},\quad S_y=\frac{1}{\sqrt{2}}\begin{bmatrix}
	0 & -\text{i} & \text{i}\\\text{i} & 0 & 0\\-\text{i} & 0 & 0
	\end{bmatrix},\quad S_x=\frac{1}{\sqrt{2}}\begin{bmatrix}
	0 & 0 & 0\\0 & -1 & 0\\0 & 0 & 1
\end{bmatrix},
\end{equation}
satisfying that
\begin{equation}
	[S_i,S_j]=\text{i}\epsilon_{ijk}S_k.
\end{equation}

\section{A 1D insulator model with CSNES and $\P\T$ symmetry}
\begin{figure}[t]
	\centering
	\includegraphics[width=4.8in]{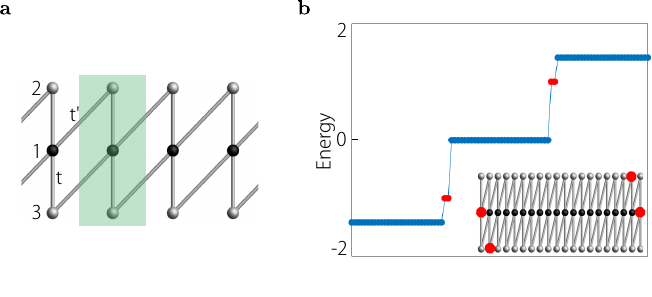}
	\caption{\textbf a presents the one-dimensional lattice structure, wherein every unit cell accommodates three sites. The amplitudes of intracell and intercell hoppings are respectively denoted as $t$ and $t'$. \textbf b shows four boundary states distributed on both ends with breaking the translational symmetry. The parameters are chosen as $t=0$ and $t'=1$.} \label{Fig:R21lattice}
\end{figure}

As shown in Fig.~\ref{Fig:R21lattice}(a), we present a one-dimensional model corresponding to the classifying space $V_{N-1}(\mathbb R^{N})$. Each unit cell consists of one $A$-site and two $B$-sites, indicated by black and grey coloring respectively. This model only incorporates hoppings between different sublattices, indicating the presence of sublattice symmetry $\S$. The intracell and intercell hopping amplitudes are denoted as $t$ and $t'$, respectively. Since all hoppings are real, the time reversal symmetry $\T$ is preserved. Furthermore, this model exhibits inversion symmetry through the center of the unit cell. The inversion symmetry $\P$ preserves the sublattices, leading to its commutation with $\S$. The combined symmetry $\P\T$ also commutes with $\S$. In momentum space, the Hamiltonian is given by
\begin{equation}\label{Eq:R21Hamil}
	\H_{\text{1D}}(k)=\begin{bmatrix}
		0&w(k)&w^*(k)\\w^*(k)&0&0\\w(k)&0&0
	\end{bmatrix},
\end{equation}
where $w_{\pm}(k)=t+t'e^{- ik}$. The symmetry operators are represented as $\S=\mathrm{diag}[1,-1_2]$ and $\mathcal{PT}=\mathrm{diag}[1,\sigma_1]\mathcal{K}$. The energy spectrum of this Hamiltonian features eigenvalues of $0$ and $\pm \sqrt{2}|w(k)|$. Notably, when $t=\pm t'$, two bulk gaps close.

To obtain the real Hamiltonian, we apply a unitary transformation $U=1\oplus e^{-i\pi/4}e^{i\pi\sigma_1/4}$ to this system. $\P\T$ symmetry is transformed as $U\P\T U^\dagger=\K$, while $\S$ is preserved. The resulting real Hamiltonian $\H_{\text{1D}}^{\mathbb R}(k)=U\H_{\text{1D}}(k)U^\dagger$ is given by
\begin{equation}
	\H^{\mathbb R}_{\text{1D}}(k)=\begin{bmatrix}
		0&w_+(k)&w_-(k)\\w_+(k)&0&0\\w_-(k)&0&0
	\end{bmatrix},
\end{equation}
where $w_{\pm}(k)=t+t'(\cos k\pm\sin k)$. For the 1D model \eqref{Eq:R21Hamil}, the expression of $\W(k)$ defined in Eq.~\eqref{eq:canonical_form} is given by 
\begin{equation}
	\mathcal W(k)=(w_+(k)\sigma_3+w_-(k)\sigma_1)/E(k).
\end{equation}
Substituting this $\mathcal W(k)$ into Eq.~\eqref{R21winding}, we find that $\nu=1$ when $|t|<|t'|$ and $\nu=0$ when $|t|>|t'|$. As depicted in Fig.~\ref{Fig:R21lattice}(b), if we break the translational symmetry and set the parameters as $t=0$ and $t'=1$, two pairs of inversion-related in-gap states emerge on the ends.

\section{The Ten-band PZSV lattice model with CSNES and $C_{2z}\T$ symmetry}

\subsection{Euler class of the flat bands}
Before presenting the model, we will first prove that the topological invariant $\mathcal N$ in Eq.~\eqref{R31winding} is actually the Euler class of the two flat bands in the bulk band structure. The corresponding system should possess chiral symmetry with $\hat \S=\text{diag}\{1_{N-2},1_N\}$, and $\mathcal{P}\mathcal{T}$ symmetry with $(\mathcal{P}\mathcal{T})^2=1$ and $[\S,\mathcal{P}\mathcal{T}]=0$. The existence of two flat bands in the bulk band structure is a consequence of the form of $\S$. The homotopy group is given by $\pi_2[V_{N-2}(\mathbb{R}^N)]\cong 2\mathbb{Z}$.

Since $\mathcal{P}\mathcal{T}$ symmetry makes $\mathcal{U}(\k)$ in Eq.~\eqref{eq:SVD} be a real orthogonal matrix, we can express $\mathcal{U}(\k)$ as 
\begin{equation}\label{eq:eigenU}
	\mathcal{U}(\k)=(
	|\psi_1(\k)\rangle,|\psi_2(\k)\rangle,\cdots,|\psi_{N-1}(\k)\rangle,|\psi_{N}(\k)\rangle),
\end{equation}
where $|\psi_i(\k)\rangle$ are real column vectors. From Eq.~\eqref{Eq:Uni-trans}, we know that 
\begin{equation}\label{eq:eigenH}
	\H(\k)|\varphi_i(\k)\rangle=0,\quad|\varphi_i(\k)\rangle=\begin{bmatrix}
		0_{(N-2)\times 1}\\|\psi_{N+i-2}(\k)\rangle
	\end{bmatrix},
\end{equation}
with $i=1$ and $2$. In other words, $|\varphi_{1,2}(\k)\rangle$ correspond to the eigenvectors of the two flat bands. Referring to Fig.~\ref{S2} and Eq.~\eqref{eq:transition}, the transition function $t(\phi)$ defined on the equator $S^1$ satisfies
\begin{equation}\label{eq:transition2}
	\begin{bmatrix} 1_{N-2} &\\&t(\phi)\end{bmatrix}=\mathcal{W}_N|_{S^1}\mathcal{W}^T_S|_{S^1}=\begin{bmatrix} \mathcal{V}_N|_{S^1} &\\&1_2\end{bmatrix}\mathcal{U}^T_N|_{S^1}\mathcal{U}_S|_{S^1}\begin{bmatrix} \mathcal{V}^T_S|_{S^1} &\\&1_2\end{bmatrix}=\mathcal{U}^T_N|_{S^1}\mathcal{U}_S|_{S^1}.
\end{equation}

Let us label the flat band eigenvectors on the semisphere $D_{\mathrm{N/S}}^2$ as $|\varphi^\mathrm{N/S}_i\rangle$. Substituting Eq.~\eqref{eq:eigenU} and Eq.~\eqref{eq:eigenH} into Eq.~\eqref{eq:transition2}, the matrix element of the $2\times 2$ matrix $t(\phi)$ is given by
\begin{equation}
	t_{ij}(\phi)=\langle \varphi^N_i(\phi)|\varphi^S_j(\phi)\rangle.
\end{equation}
The above equation shows that the transition function $t(\phi)$ of $\W(\k)$ is also the transition function of the eigenvectors of the flat bands. Moreover, we can apply the Wilson loop method to the flat bands, and the Wilson loop operator has the same eigenvalues as $t(\phi)$. From the definition of the Euler class, we know that the winding number $\mathcal N$ in Eq.~\eqref{R31winding} corresponds to the Euler class of the two flat bands. Thus, the topological invariant $\mathcal N$ is equal to the number of times that the Wilson loop spectra crosses $\pi$.

\subsection{The PZSV model}
Near the first magic angle ($\theta\approx 1.05^{\circ}$), twisted bilayer graphene (TBG) exhibits two almost perfectly flat bands for each spin and valley degree of freedom. In this work, we adopt the ten-band Po-Zou-Senthil-Vishwanath (PZSV) tight-binding model proposed by Vishwanath to capture the low-energy physics of single-valley TBG (See Ref. [12] in the main text). This model preserves both the CSNES and $C_{2z}\T$ symmetry. In two dimensions, $C_{2z}\T$ symmetry imposes the same constraints on $\H(\k)$ as $\P\T$ symmetry. And the two nearly flat bands in the bulk band structure of TBG arise from the CSNES.

\begin{figure}[t]
	\includegraphics[width=5.2in]{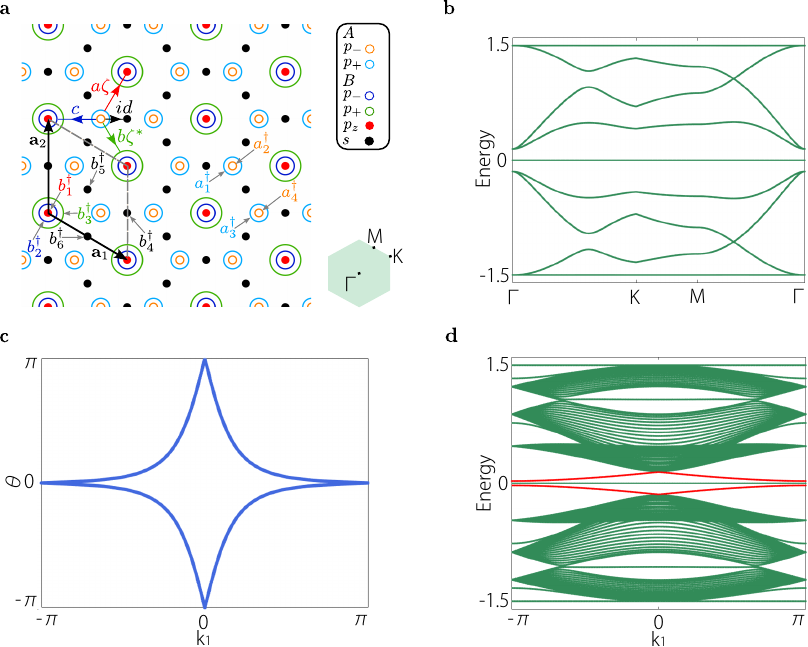}
	\caption{\textbf a illustrates the lattice structure of the ten-band PZSV lattice model. The unit cell, represented by a dashed gray square, consists of 10 orbitals. As detailed in the inset at the top-right of the panel, $A$-sublattice contains two $p_-$-orbitals and two $p_+$-orbitals, whereas $B$-sublattice contains one $p_-$-orbital, one $p_+$-orbital, one $p_z$-orbital and three $s$-orbitals. Here, we use the convention of Ref. [12] in the main text, where $p_\pm=\frac{1}{\sqrt 2} (p_x\pm ip_y)$. The first Brilliouin zone is shown in the bottom-right inset. In \textbf b, \textbf c and \textbf d, the hopping parameters are set as $a=0.110$, $b=0.033$, $c=0.033$, and $d=0.573$. \textbf b displays the bulk band structure of the PZSV model, where two flat bands exist because of the form of CSNES. \textbf c shows the Wilson loop spectra of the flat bands along the circles $C_{k_1}$ ($k_1$ is fixed for each circle) parameterized by $k_2$. \textbf d shows the egde states when the boundary along $\mathbf a_1$ is opened.}
	\label{FIG:PZSV}
\end{figure}

The lattice structure of the PZSV model is illustrated in Fig.~\ref{FIG:PZSV}(a), where each unit cell contains ten orbitals: $a^\dagger_i$ with $i$ ranging from $1$ to $4$ for $A$-sites, and $b^\dagger_j$ with $j$ ranging from $1$ to $6$ for $B$-sites. The specific orbital content of each sublattice is shown in the top-right inset of Fig.~\ref{FIG:PZSV}(a). The hopping terms only exist between $a^\dagger_i$ and $b^\dagger_j$, and thus the chiral symmetry can be expressed as
\begin{equation}
	\SS=\begin{bmatrix}
		1_4 &\\&-1_6
	\end{bmatrix}.
\end{equation}
The PZSV model is invariant under $C_{2z}\T$ symmetry with the inversion center located at the blue dot labeled by $b_1^\dagger$ in Fig.~\ref{FIG:PZSV}(a). Under the $C_{2z}\T$ operation, $s$-orbitals and $p_z$-orbitals remain invariant, while the $p_+$-orbitals and $p_-$-orbitals are exchanged. Therefore, the $C_{2z}\T$ symmetry is given by
\begin{equation}
	\hat{C}_{2z}\TT=\begin{bmatrix}
		0 & 0 & 0 & 1 & 0 & 0 & 0 & 0 & 0 & 0\\
		0 & 0 & 1 & 0 & 0 & 0 & 0 & 0 & 0 & 0\\
		0 & 1 & 0 & 0 & 0 & 0 & 0 & 0 & 0 & 0\\
		1 & 0 & 0 & 0 & 0 & 0 & 0 & 0 & 0 & 0\\
		0 & 0 & 0 & 0 & 1 & 0 & 0 & 0 & 0 & 0\\
		0 & 0 & 0 & 0 & 0 & 0 & 1 & 0 & 0 & 0\\
		0 & 0 & 0 & 0 & 0 & 1 & 0 & 0 & 0 & 0\\
		0 & 0 & 0 & 0 & 0 & 0 & 0 & 1 & 0 & 0\\
		0 & 0 & 0 & 0 & 0 & 0 & 0 & 0 & 1 & 0\\
		0 & 0 & 0 & 0 & 0 & 0 & 0 & 0 & 0 & 1\\
	\end{bmatrix}\K.
\end{equation}

Since $\hat \S=\text{diag}\{1_4,-1_6\}$, $(C_{2z}\T)^2=1$, and $[\S,\C_{2z}\T]=0$, the PZSV model is classified into the real Stiefel manifold $V_4(\mathbb R^6)$. The Bravais lattice vectors of the PZSV lattice are chosen as
\begin{equation}
	\mathbf{a}_1=(\sqrt 3/2,-1/2,0),~\mathbf a_2=(0,1,0),~\mathbf a_3=(0,0,1).
\end{equation}
The translation along the $z$-direction is only considered for calculating the reciprocal lattice vectors $\mathbf b_{1,2}$, which are given by
\begin{equation}
	\mathbf{b}_1=\frac{\mathbf a_2\times\mathbf a_3}{\mathbf a_1\cdot(\mathbf a_2\times\mathbf a_3)}=(\frac{2}{\sqrt 3},0,0),~\mathbf{b}_2=\frac{\mathbf a_3\times\mathbf a_1}{\mathbf a_2\cdot(\mathbf a_3\times\mathbf a_1)}=(\frac{1}{\sqrt 3},1,0).
\end{equation}
Then, in the momentum space, the Fourier-transformed operators are given by
\begin{equation}
	a^\dagger_{\k,i}=\frac{1}{\sqrt N}\sum_{\mathbf R}a^\dagger_{\mathbf R,i}e^{-\text{i}\k\cdot(\mathbf R+\mathbf r_i^{a})},~	b^\dagger_{\k,j}=\frac{1}{\sqrt N}\sum_{\mathbf R}b^\dagger_{\mathbf R,j}e^{-\text{i}\k\cdot(\mathbf R+\mathbf r_j^{b})},
\end{equation}
where $\mathbf R$ labels the unit cell, $\mathbf r_i^{a/b}$ denotes the displacement of the $a^\dagger_{\mathbf R,i}$ fermion relative to the unit cell origin, and $\k=k_1\mathbf b_1+k_2\mathbf b_2$ represents the crystalline momentum. Thus, the Bloch Hamiltonian is given by
\begin{equation}
	H(\k)=\sum_{\k}\Psi^\dagger(\k) \H(\k) \Psi(\k),
\end{equation}
with the momentum space spinor
\begin{equation}
	\Psi^\dagger(\k)=(a^\dagger_{\k,1},\cdots,a^\dagger_{\k,4},b^\dagger_{\k,1},\cdots,b^\dagger_{\k,6}),
\end{equation}
and the first-quantized Hamiltonian
\begin{equation}\label{eq:PZSV}
	\H(\k)=\begin{bmatrix}
		&\Q(\k)\\\Q^\dagger(\k) &
	\end{bmatrix}.
\end{equation}
Here, the elements of the $4\times 6$ matrix $\Q(\k)$ are obtained through the Fourier transformation
\begin{equation}
	[\Q(\k)]_{ij}=\sum_{\mathbf R}[\Q_{\mathbf R}]_{ij}e^{\text{i}\k\cdot(\mathbf{R}+\mathbf{r}_i^a-\mathbf r_j^b)},
\end{equation}
where $\Q_{\mathbf R}$ represents the inter-sublattice hopping matrix. The $i$-th column of the $4\times 6$ matrix $\Q(\k)$, denoted by $\Q_{:,i}(\k)$, can be expressed as
\begin{equation}
		\Q_{:,1}=\begin{bmatrix}
			-a \zeta e^{ i (k_1 + 2k_2)/3} (\eta e^{-i (k_1+k_2)} + e^{-i k_2} + \eta^*)\\
			a \zeta e^{i(k_1 + 2k_2)/3}(\eta e^{-i (k_1+k_2)} + \eta^*e^{-ik_2}+1))\\
			a \zeta^* e^{i(2k_1 + k_2)/3}(e^{-i (k_1+k_2)} + \eta e^{-ik_1} + \eta^*)\\
			-a \zeta^* e^{i(4k_1 + 5k_2)/6}(\eta e^{-i(k_1 + 3k_2/2)}+ e^{-i(k_1+k_2/2)} + \eta^* e^{-ik_2/2})
		\end{bmatrix},
\end{equation}
\begin{equation}
		\Q_{:,2}=\begin{bmatrix}
			b\zeta^* e^{ i (k_1 + 2k_2)/3} (\eta +\eta^* e^{-i (k_1+k_2)}+ e^{-i k_2})\\
			c e^{ i (k_1 + 2k_2)/3}(e^{-i(k_1 + k_2)} + e^{-ik_2} + 1)\\
			b\zeta e^{i(4k_1 + 5k_2)/6}(e^{-i(k_1 + 3k_2/2)} + \eta^* e^{-i(k_1+k_2/2)} + \eta e^{-ik_2/2})\\
			c (e^{-i(k_1 + 2k_2)/3} + e^{i(-k_1 + k_2)/3} + e^{i(2k_1 + k_2)/3})
		\end{bmatrix},
	\end{equation}
	\begin{equation}
		\Q_{:,3}=\begin{bmatrix}
			c e^{ i (k_1 + 2k_2)/3}(e^{-i(k_1 + k_2)} + e^{-ik_2} + 1)\\
			b\zeta^* e^{ i (k_1 + 2k_2)/3} (1 +\eta^* e^{-i (k_1+k_2)}+ \eta e^{-i k_2})\\
			c (e^{-i(k_1 + 2k_2)/3} + e^{i(-k_1 + k_2)/3} + e^{i(2k_1 + k_2)/3})\\
			b\zeta e^{i(4k_1 + 5k_2)/6}(\eta^*e^{-i(k_1 + 3k_2/2)} +  e^{-i(k_1+k_2/2)} + \eta e^{-ik_2/2})
		\end{bmatrix},
	\end{equation}
	\begin{equation}
		\Q_{:,4}=\begin{bmatrix}
			i d e^{i (2 k_1 + k_2)/6}\\
			i d e^{i (2 k_1 + k_2)/6}\\
			-i d e^{-i(2k_1 + k_2)/6} \\
			-i d e^{-i(2k_1 + k_2)/6}
		\end{bmatrix},~\Q_{:,5}=\begin{bmatrix}
			i d\eta^*  e^{ i (-k_1 + k_2)/6}\\
			i d\eta  e^{ i (-k_1 + k_2)/6}\\
			-i d\eta^*  e^{ i (k_1 - k_2)/6}\\
			-i d\eta  e^{ i (k_1 - k_2)/6}\\
		\end{bmatrix},~\Q_{:,6}=\begin{bmatrix}
			i d\eta  e^{ -i (k_1 + 2k_2)/6}\\
			i d\eta^*  e^{ -i (k_1 + 2k_2)/6}\\
			i d\eta  e^{ i (k_1 + 2k_2)/6}\\
			i d\eta^*  e^{ i (k_1 + 2k_2)/6}\\
		\end{bmatrix}.
\end{equation}
Here, $a$, $b$, $c$, and $d$ are hopping parameters, while $\zeta=e^{i\pi/3}$ and $\eta=\zeta^2$ are fixed values.

In the first Brillouin zone $(k_1,k_2)\in[-\pi,\pi]^2$, the Hamiltonian $\H(\k)$ is non-periodic. However, we can consider the periodic Hamiltonian $\H'(\k)=V(\k)\H(\k)V^\dagger(\k)$, where $V(\k)$ is given by
\begin{equation}
	V(\k)=\begin{bmatrix}
		e^{2i(k_1-k_2)/3} & 0 & 0 & 0 & 0 & 0 & 0 & 0 & 0 & 0\\
		0 & e^{2i(k_1-k_2)/3} & 0 & 0 & 0 & 0 & 0 & 0 & 0 & 0\\
		0 & 0 & e^{i(k_1-k_2)/3} & 0 & 0 & 0 & 0 & 0 & 0 & 0\\
		0 & 0 & 0 & e^{i(k_1-k_2)/3} & 0 & 0 & 0 & 0 & 0 & 0\\
		0 & 0 & 0 & 0 & 1 & 0 & 0 & 0 & 0 & 0\\
		0 & 0 & 0 & 0 & 0 & 1 & 0 & 0 & 0 & 0\\
		0 & 0 & 0 & 0 & 0 & 0 & 1 & 0 & 0 & 0\\
		0 & 0 & 0 & 0 & 0 & 0 & 0 & e^{-ik_2/2} & 0 & 0\\
		0 & 0 & 0 & 0 & 0 & 0 & 0 & 0 & e^{i(k_1-k_2)/2} & 0\\
		0 & 0 & 0 & 0 & 0 & 0 & 0 & 0 & 0 & e^{ik_1/2}\\
	\end{bmatrix}.
\end{equation}
The matrix element of $V(\k)$ is obtained as $[V(k)]_{ij}=\delta_{ij}e^{\text i \k \cdot\mathbf r_i}$, where $\mathbf r_i$ denotes the displacement of the $a^\dagger_{\mathbf R,i}$ ($b^\dagger_{\mathbf R,i-4}$) fermion relative to the unit cell origin for $1\leq i\leq 4$ ($5\leq i\leq 10$).

For a reciprocal lattice translation $\mathbf K=2\pi(m\mathbf b_1+n\mathbf b_2)$ with $m,n\in\Z$, we have 
\begin{equation}
	\H(\k+\textbf{K})=V^\dagger(\textbf K)\H(\k)V(\textbf K).
\end{equation} 
Thus, given the eigenvectors $|\psi_i(\k)\rangle$ of $\H(\k)$, $V^\dagger(\textbf K)|\psi_i(\k)\rangle$ should be the eigenvectors of $\H(\k+\mathbf K)$. Now, to elaborate the Wilson loop calculation for the non-periodic Hamiltonian $\H(\k)$ in Eq.~\eqref{eq:PZSV}, we consider the 1D insulator subsystem and divide the 1D first Brillouin zone into $N$ intervals labeled by separation points $i = 0,1,\cdots,N$. The Wilson loop operator $W^{(\Lambda)}$ is then expressed as
\begin{equation}
	W^{(\Lambda)}=\lim_{N\rightarrow\infty}\big{(}\prod_{i=0}^{N-2}F_{(i,i+1)}\big{)}\times F^{(\Lambda)}_{(N-1,N)},
\end{equation}
with 
\begin{equation}
	[F_{(i,i+1)}]_{ab}=\langle a,k_{(i)}|b,k_{(i+1)}\rangle,~[F^{(\Lambda)}_{(N-1,N)}]_{ab}=\langle a,k_{(N-1)}|\Lambda|b,k_{(0)}\rangle.
\end{equation}
Here, the indices $a$ and $b$ label the bands, and $\Lambda=V^\dagger(2\pi\mathbf b_{1,2})$ is used when calculating the Wilson loop along the circles $C_{k_{2,1}}$ (with $k_{2,1}$ fixed for each circle) parameterized by $k_{1,2}$, respectively.

In Vishvanath's paper, the PZSV model is characterized by four hopping parameters: $a=0.110$, $b=0.033$, $c=0.033$, and $d=0.573$ (in dimensionless units). Under this parameter condition, the bulk band structure is shown in Fig.~\ref{FIG:PZSV}(b). The form of $\S$ induces two flat bands. The Wilson loop spectra of the two flat bands are displayed in Fig.~\ref{FIG:PZSV}(c), revealing an Euler class of $1$. Therefore, the topological invariant defined in Eq.~\eqref{R31winding} is $\mathcal N=1$. Notably, $\mathcal N$ is not an even number due to the non-periodicity of $\H(\k)$. If we consider the Hamiltonian \eqref{eq:PZSV} within the extended period $(k_1,k_2)\in[-6\pi,6\pi]^2$, $\mathcal N$ would be the even number $36$. The egde states with opening the boundary along $\mathbf a_1$ are shown in Fig.~\ref{FIG:PZSV}(d).